\documentclass[11pt,a4paper]{article}
\pdfoutput=1

\usepackage{jheppub}
\usepackage{amsmath}
\usepackage{amssymb}
\usepackage{graphics}
\usepackage[active]{srcltx}
\usepackage{pdfsync}
\usepackage{shuffle}
\usepackage{slashed}
\usepackage{hyperref}
\usepackage{subfigure}

\newcommand{\bea}{\begin{eqnarray}}
\newcommand{\eea}{\end{eqnarray}}
\newcommand{\nn}{\nonumber \\}

\def\eref#1{(\ref{#1})}

\def\a{{\alpha}}

\allowdisplaybreaks

\title{Recursion Relation for BCFW Boundary Terms of NLSM Amplitudes}
\author[a]{Kang Zhou}

\affiliation[a]{Center for Gravitation and Cosmology, College of Physical Science and Technology, Yangzhou University,\\
No.180, Siwangting Road, Yangzhou, 225009, P.R. China}

\emailAdd{zhoukang@yzu.edu.cn}
\date{\today}
\abstract{When computing tree-level amplitudes of the nonlinear sigma model (NLSM) via the BCFW on-shell recursion relation, boundary terms are inevitable. Here, the BCFW recursion relation refers to the original prescription with momentum-shifts on two external legs, as opposed to the generalized versions with redefined momentum-shifts. We propose a simple general scheme that determines the boundary terms of NLSM amplitudes recursively from lower-point amplitudes. Hidden zeros and mass dimensions severely constrain boundary terms, allowing us to compute these terms from the $2$-split property of the tree amplitudes. Combined with the existing rules in the literature for obtaining the two currents in the $2$-split from on-shell amplitudes, this scheme enables the recursive construction of boundary terms for higher-point amplitudes from lower-point amplitudes.
}


\begin{document}

\maketitle \flushbottom

%%%%%%%%%%%%%%%%%%%%%%%%%%%%%%
\section{Introduction}
\label{sec-intro}
%%%%%%%%%%%%%%%%%%%%%%%%%%%%%%%%

In modern studies of scattering amplitudes, on-shell methods occupy an important position. On-shell methods avoid the unphysical degrees of freedom carried by off-shell virtual particles, leading to a tremendous improvement in computational efficiency. At tree-level, the representative of on-shell methods is the well-known Britto-Cachazo-Feng-Witten (BCFW) on-shell recursion relation \cite{Britto:2004ap,Britto:2005fq}. This method is based on locality and unitarity, assembling higher-point on-shell amplitudes from lower-point ones. The BCFW on-shell recursion relation not only brought about a revolutionary breakthrough in computation, but also stimulated a great deal of research into new structures and new formalisms of scattering amplitudes \cite{Arkani-Hamed:2009ljj,Arkani-Hamed:2012zlh,Arkani-Hamed:2013jha}.

The BCFW on-shell recursion relation has suffered from the so-called boundary term problem since its inception. When applied to Yang-Mills and gravitational amplitudes at the tree-level, it can be shown that the suitable BCFW-shifts can be found such that the boundary terms vanish \cite{Arkani-Hamed:2008bsc}. In this case, the BCFW method exhibits a tremendous advantage in computational efficiency. However, when attempting to apply this method to amplitudes in other physical models, such as those of certain low-energy effective theories, the boundary terms become the greatest obstacle. Therefore, the boundary term difficulty is the primary bottleneck restricting the range of applicability of the BCFW method.

The existence of boundary terms indicates that locality and unitarity, which serve as the foundation of the BCFW method, are in many cases insufficient to fully determine the amplitudes. Therefore, to resolve the boundary term difficulty, it is foreseeable that one needs to resort to physical properties beyond locality and unitarity. The recently discovered hidden zeros of tree amplitudes and the novel factorization behavior near them, called $2$-split, are excellent candidates for such new properties \cite{Arkani-Hamed:2023swr,Cao:2024gln,Cao:2024qpp}.
A series of related studies have shown that combining locality and unitarity with hidden zeros or $2$-splits can completely determine the corresponding tree-level amplitudes \cite{Rodina:2024yfc,Zhou:2025xly,Zhou:2026isc}. These results provide a foundation for using hidden zeros and $2$-splits to address the boundary term problem in the BCFW method.

The tree amplitudes currently known to exhibit hidden zeros encompass the following models \cite{Bartsch:2024amu,Li:2024qfp,Zhang:2024efe,Huang:2025blb,Zhou:2025tvq,CarrilloGonzalez:2026lnu,Zhang:2026wye}: ${\rm Tr}(\phi^3)$, nonlinear sigma model (NLSM), Yang-Mills (YM), Dirac-Born-Infeld, gravity (GR), special Galileon, certain couplings among the above models, YM and GR amplitudes with specific higher-derivative interactions, as well as specific massive models and specific Fermionic amplitudes. Most of the above amplitudes satisfy the $2$-splits (or the generalized $2$-splits) near the hidden zeros \cite{Cao:2024gln,Cao:2024qpp,Zhang:2026dcm}.
If the boundary term difficulty can be resolved through hidden zeros and the $2$-splits near them, then the range of applicability of the BCFW on-shell recursion relation would be extended to include the aforementioned amplitudes\footnote{Several form factors also satisfy the hidden zeros\cite{Cao:2025hio,Zhang:2025zjx}, while several string theory amplitudes also satisfy the $2$-split\cite{Cao:2024gln,Cao:2024qpp}. This paper only concerns field theory amplitudes.}.

Motivated by the above expectation, we recently modified the BCFW on-shell recursion relation by exploiting hidden zeros, rendering it capable of computing NLSM tree amplitudes \cite{Li:2025suo}. By altering the integrand of the contour integral in the usual BCFW method, we made the boundary terms of the NLSM amplitudes vanish, while the potential new poles introduced by the new integrand are eliminated by the hidden zeros, thereby yielding a recursion relation free of boundary terms. However, in our approach the shift of kinematic variables is rather complicated, resulting in a large number of poles whose residues must be computed; whereas the original BCFW method shifts the momenta of only two external legs, requiring far fewer residues to be evaluated (this point will be explained in section \ref{subsec-boundary}). Consequently, although our new method overcomes the boundary term problem, it sacrifices part of the computational efficiency advantage of the BCFW method.

In view of the above shortcoming, in this paper we attempt to retain the shift of external momenta in the original BCFW method, thereby minimizing the number of poles whose residues need to be computed, while developing a method for computing the boundary terms by means of hidden zeros and the $2$-splits near the zeros. That is to say, we do not intend to eliminate the boundary terms; instead, we plan to construct them. In section \ref{subsec-0-and-split}, we will show that, from the perspective of Feynman diagrams, there exists a striking complementary relation between the factorization of amplitudes at the poles used in the BCFW method and the factorization called $2$-split near the hidden zeros. This complementary relation suggests that the $2$-split may compensate for what is missing in the usual BCFW method. Therefore, the $2$-split plays an important role in the present work.

In this work, we still focus mainly on NLSM tree amplitudes, since they are typical effective theory amplitudes for which the boundary terms can by no means be avoided within the standard BCFW method.
We combine the hidden zeros with an analysis of the general properties of boundary terms and of the mass dimensions of the target amplitudes, in order to derive the general constraints that the boundary terms of the target amplitudes should satisfy. Here the target amplitudes include, in addition to the NLSM amplitudes, NLSM$\oplus{\rm Tr}(\phi^3)$ amplitudes involving three external ${\rm Tr}(\phi^3)$ scalar particles. Our analysis shows that the NLSM$\oplus{\rm Tr}(\phi^3)$ amplitudes have no boundary terms when the shifted momenta are those of the ${\rm Tr}(\phi^3)$ scalar external legs; on the other hand, once appropriate independent Mandelstam variables are chosen, the boundary terms of the NLSM amplitudes satisfy a simple expansion form, with the expansion coefficients obeying specific constraints.
We then find that these expansion coefficients can be systematically probed by a series of $2$-splits. Meanwhile, the work in \cite{Azevedo:2025vxo} allows us to obtain the expressions for the off-shell currents in the $2$-split from those for the on-shell amplitudes.
Combining these two observations, we obtain a simple formula \eref{recursionform-boundary} for computing the boundary terms of NLSM amplitudes.
This formula recursively constructs the boundary terms of higher-point NLSM amplitudes from lower-point on-shell NLSM$\oplus{\rm Tr}(\phi^3)$ and NLSM amplitudes, in full accord with the on-shell recursive spirit of the BCFW method.

Naturally, our ultimate goal is to compute as many boundary terms as possible for amplitudes satisfying the hidden zeros and $2$-splits. Therefore, this paper constitutes an initial step toward this endeavor.

The remainder of this paper is organized as follows. In section \ref{sec-background}, we review the necessary background. In section \ref{sec-fix-boundary}, we discuss how hidden zeros, combined with other physical conditions, constrain the behavior of the boundary terms. We will show that the NLSM$\oplus{\rm Tr}(\phi^3)$ amplitudes of interest in this paper have no boundary terms, and construct a general method for recursively computing the boundary terms of NLSM amplitudes from lower-point amplitudes. In section \ref{sec-example-nlsm+phi}, we compute two examples of $5$-point NLSM$\oplus{\rm Tr}(\phi^3)$ amplitudes via the standard BCFW method to verify the conclusion, obtained in section \ref{sec-fix-boundary}, that the boundary terms of these amplitudes vanish. In section \ref{sec-example-nlsm}, we employ the general method developed in section \ref{sec-fix-boundary} to compute the boundary terms of the same $6$-point NLSM amplitude for two distinct BCFW-shifts, and combine them with the recursively computable parts given by the standard BCFW method to obtain the correct $6$-point amplitude, thereby validating the method for computing boundary terms proposed in this paper. Finally, we present a brief summary and discussion in section \ref{sec-conclu}.

%%%%%%%%%%%%%%%%%%%%%%%%%%%%%%
\section{Background}
\label{sec-background}
%%%%%%%%%%%%%%%%%%%%%%%%%%%%%

For readers' convenience, this section introduces some background material used in this paper. In subsection \ref{subsec-boundary}, we briefly review the BCFW on-shell recursion relation and the issue of boundary terms therein. In subsection \ref{subsec-0-and-split}, we introduce the hidden zeros of tree-level amplitudes---particularly those of the NLSM---as well as the novel factorization behavior near these zeros, known as $2$-split. In subsection \ref{subsec-onshell-current}, we describe how to obtain the two off-shell currents in $2$-split starting from the expressions for the on-shell amplitudes.

%%%%%%%%%%%%%%%%%%%%%%%%%%%%%%
\subsection{BCFW recursion relation and boundary term}
\label{subsec-boundary}
%%%%%%%%%%%%%%%%%%%%%%%%%%%%%%

The basic idea of the BCFW recursion relation  is to promote a tree-level amplitude to a function ${\cal A}(z)$ of a single complex variable
$z$ via an appropriate shift of external momenta, and to extract the residue at
$z=0$, which corresponds to the desired physical amplitude, from the residues at $z\neq0$. Since locality and unitarity guarantee that the residues at
$z\neq0,\infty$ factorize into products of two lower-point amplitudes, this yields a recursive construction of higher-point on-shell amplitudes by gluing together lower-point on-shell amplitudes.

The deformation of external momenta must preserve both momentum conservation and the on-shell conditions. The simplest option is to choose two external legs, $i$ and $j$, and deform their momenta as
\bea
\hat{k}_i=k_i+zq\,,~~~~\hat{k}_j=k_j-zq\,,~~~~~~~~{\rm where}~
k_i\cdot q=k_j\cdot q=q^2=0\,.~~~~\label{BCFW-shift}
\eea
The on-shell $n$-point tree amplitude ${\cal A}_n(z)$, which depends on $\hat{k}_i(z)$ and $\hat{k}_j(z)$, is a function of
$z$. Consider the following contour integral,
\bea
{1\over 2\pi i}\,\oint\,{dz\over z}\,{\cal A}_n(z)\,.~~~~\label{BCFW-countorintegral}
\eea
The residue of this integral around $z=0$ gives ${\cal A}_n(0)$, which is precisely the desired amplitude prior to the BCFW-shift in \eref{BCFW-shift}.
The residue theorem tells us that
\bea
{\cal A}_n(0)&=&{\rm Res}_{z=0}\,{{\cal A}_n(z)\over z}\nn
&=&-\sum_{z^*\neq0,\infty}\,{\rm Res}_{z=z^*}\,{{\cal A}_n(z)\over z}-{\rm Res}_{z=\infty}\,{{\cal A}_n(z)\over z}\,.
\eea
Locality and unitarity imply that each pole at $z^*\neq0,\infty$ arises from an internal propagator going on-shell, and that the residue of ${\cal A}_n(z)$ at each $z^*$ factorizes into a product of two lower-point on-shell amplitudes. This observation yields
\bea
{\cal A}_n(0)
&=&\sum_p\,\sum_{\pmb I}\,{\cal A}_{n_L}(z^*_p)\,{1\over D_p(0)}\,{\cal A}_{n_R}(z^*_p)\Big|_{D_p(z^*_p)=0}-{\rm Res}_{z=\infty}\,{{\cal A}_n(z)\over z}\,,~~~~~~~~{\rm where}~
n_L+n_R=n+2\,.~~~~\label{BCFW-recursive-form}
\eea
In the above, each $D_p(z)$ is the denominator of the internal propagator depending on
$z$, while each $D_p(0)$ is the corresponding denominator prior to the BCFW-shift. In other words, $D_p(z)$ correspond to the propagators that are detectable by the BCFW-shift \eref{BCFW-shift}.
The summation $\sum_{\pmb I}$ denotes the sum over the possible internal degrees of freedom of the on-shell virtual particle, for example, the sum over the helicities of the virtual gluon.

When the residue at infinity vanishes, namely ${\rm Res}_{z=\infty}=0$, the expression \eref{BCFW-recursive-form} provides a perfect realization of the recursive construction from lower-point to higher-point amplitudes. It is worth noting that the deformation of external momenta in \eref{BCFW-shift} is not only the simplest in form, but also the most economical in computation. This is because, under the BCFW-shift of \eref{BCFW-shift}, only the propagators lying on the line $L_{(i,j)}$ connecting the external legs $i$ and $j$ depend on
$z$; consequently, only the residues arising from these propagators going on shell need to be computed. When more external momenta are deformed, more propagators depend on $z$, and the number of residues to be evaluated grows rapidly with the number of deformed external momenta. This is why, as stated in section \ref{sec-intro}, we favour the original deformation involving a shift of only two external momenta, and seek to avoid various generalized deformations that shift many external momenta.

The residue ${\rm Res}_{z=\infty}$ at infinity is referred to as the boundary term. For tree amplitudes in many physical models, such as NLSM amplitudes, the boundary terms are unavoidable. How to compute the boundary terms in NLSM amplitudes is the subject of this paper. We therefore proceed to analyse some properties of the boundary term, which will play an important role in the subsequent sections.
Since ${\cal A}_n(z)$ contains only simple poles, one can expand ${\cal A}_n(z)$ as
\bea
{\cal A}_n(z)=\sum_p\,{R_p(z^*_p)\big|_{D_p(z^*_p)=0}\over D_p(z)}\,+\,C_0\,+\,\sum_{m\geq1}\,C_m z^m\,.~~~~\label{expan-A(z)}
\eea
The physical amplitude corresponds to $z=0$, at which all $z^m$ vanish, leaving only
\bea
{\cal A}_n\,=\,\sum_p\,{R_p(z^*_p)\big|_{D_p(z^*_p)=0}\over D_p(0)}\,+\,C_0\,.
\eea
The first part at the r.h.s. can be computed recursively via \eref{BCFW-recursive-form}, while the second part is the boundary term.
From this we observe two features of the boundary term. First, it is the constant term in the expansion of ${\cal A}_n(z)$ in
$z$. Second, since the boundary term contains no poles, it also contains no propagators that can be probed by the shift in \eref{BCFW-shift}. That is, the boundary term contains no effective propagators along the line $L_{(i,j)}$. In section \ref{subsec-graph-boundary}, we will analyse the second characteristic in more detail for scalar amplitudes.

%%%%%%%%%%%%%%%%%%%%%%%%%%%%%%
\subsection{Hidden zeros and $2$-splits of tree amplitudes}
\label{subsec-0-and-split}
%%%%%%%%%%%%%%%%%%%%%%%%%%%%%%

Hidden zeros refer to the phenomenon that tree-level amplitudes in many physical models vanish when the external momenta satisfy certain constraints. The procedure for realizing the zeros is as follows. For an
$n$-point amplitude, one first selects two external legs $i$ and
$j$, and then divides the remaining external legs into two sets
$A$ and $B$, i.e. $A\cup B=\{1,\cdots,n\}\setminus\{i,j\}$. For ordered amplitudes, there exists a natural choice of
$A$ and $B$ (both of which are ordered), namely $A=\{i+1,\cdots,j-1\}$ and $B=\{j+1,\cdots,i-1\}$, where the ordering is assumed to be $(1,2,\cdots,n-1,n)$. For NLSM amplitudes and NLSM$\oplus{\rm Tr}(\phi^3)$ amplitudes involving three ${\rm Tr}(\phi^3)$ scalar particles, which will be the focus of this paper, each amplitude vanishes when the external momenta satisfy
\bea
k_a\cdot k_b=0\,,~~~~{\rm for}~\forall\,a\in A\,,~b\in B\,,~~~~A\cup B=\{1,\cdots,n\}\setminus\{i,j\}\,,~~~~\label{kine-condi-0-scalar}
\eea
that is,
\bea
{\cal A}_n\,&\xrightarrow[]{\eref{kine-condi-0-scalar}}&\,0\,.~~\label{zero-scalar}
\eea
For gluon or graviton amplitudes, the zero conditions also include analogous restrictions on the polarization vectors or polarization tensors of the external particles.

\begin{figure}
  \centering
   \includegraphics[width=16cm]{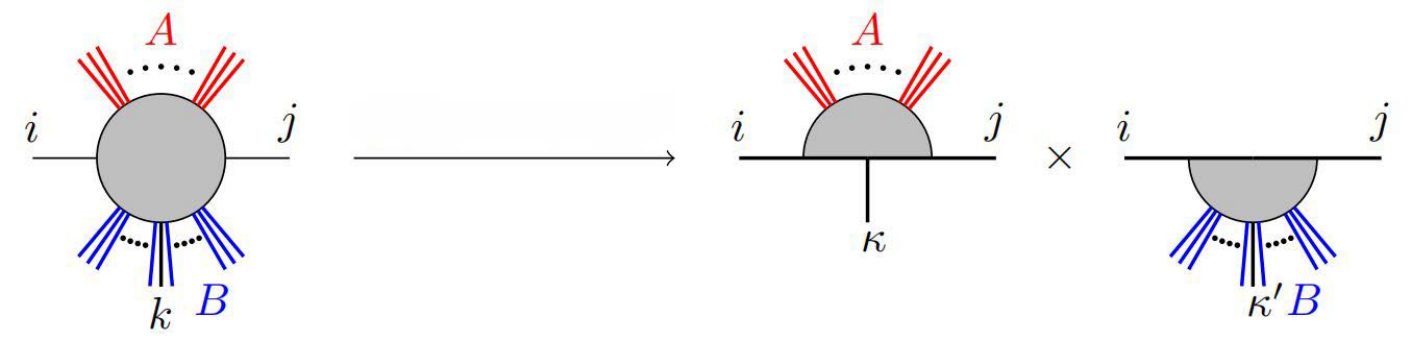} \\
  \caption{The novel factorization behavior near hidden zero, known as $2$-split.}\label{graph-2split}
\end{figure}

By removing an external leg $k$ from the zero condition \eref{kine-condi-0-scalar}, one obtains a new factorization behavior near the zero, which is called the $2$-split. Without loss of generality, we assume that $k\in B$. Then, for scalar amplitudes, the kinematic condition for $2$-split is that $k$ is
\bea
k_a\cdot k_b=0\,,~~~~{\rm for}~\forall\,a\in A\,,~b\in B\setminus k\,,~~~~A\cup B=\{1,\cdots,n\}\setminus\{i,j\}\,.~~~~\label{kine-condi-split-scalar}
\eea
Here we present only the $2$-split of the tree-level NLSM amplitudes that will be used in this paper: the $n$-point NLSM amplitude ($n$ is even) factorizes into the following two currents,
\bea
{\cal A}^{\rm NLSM}_n(1,\cdots,n)\,&\xrightarrow[]{\eref{kine-condi-split-scalar}}&\,{\cal J}^{\rm NLSM}_{n_1}(i,A,j,\kappa)\,\times\,{\cal J}^{{\rm NLSM}\oplus{\rm Tr}(\phi^3)}_{n_2}(i_\phi,j_\phi,B(\kappa'_\phi))\,,\nn
&&{\rm where}~n_1+n_2=n+3.~~\label{2split-scalar}
\eea
In \eref{2split-scalar}, $B(\kappa'_\phi)$ denotes the replacement of $k$ in the ordered set $B$ by $\kappa'$. In the two currents, the off-shell external legs $\kappa$ and $\kappa'$ can be regarded as inherited from $k$, as depicted in Fig.\ref{graph-2split}, but momentum conservation forces them to carry the following off-shell momenta,
\bea
k_\kappa&=&\sum_{b\in B}\,k_b\,,~~~~~~~~k_{\kappa'}=k_k+\sum_{a\in A}\,k_a\,.
\eea
We have assumed that $n_1$ is even and $n_2$ is odd, in which case the external legs $i,j,\kappa'$ in the current ${\cal J}_{n_2}$ are interpreted as the particles of ${\rm Tr}(\phi^3)$ model, as labeled by the subscript $\phi$. All remaining particles, whether in ${\cal J}_{n_1}$ or in ${\cal J}_{n_2}$, are interpreted as pions. Therefore, ${\cal J}_{n_2}$ is a pure NLSM current. If $n_1$ is odd and $n_2$ is even, then the external legs $i,j,\kappa$ in the current ${\cal J}_{n_1}$ are interpreted as the particles of ${\rm Tr}(\phi^3)$, while the current ${\cal J}_{n_2}$ is understood as a pure NLSM current.

\begin{figure}
  \centering
   \includegraphics[width=12cm]{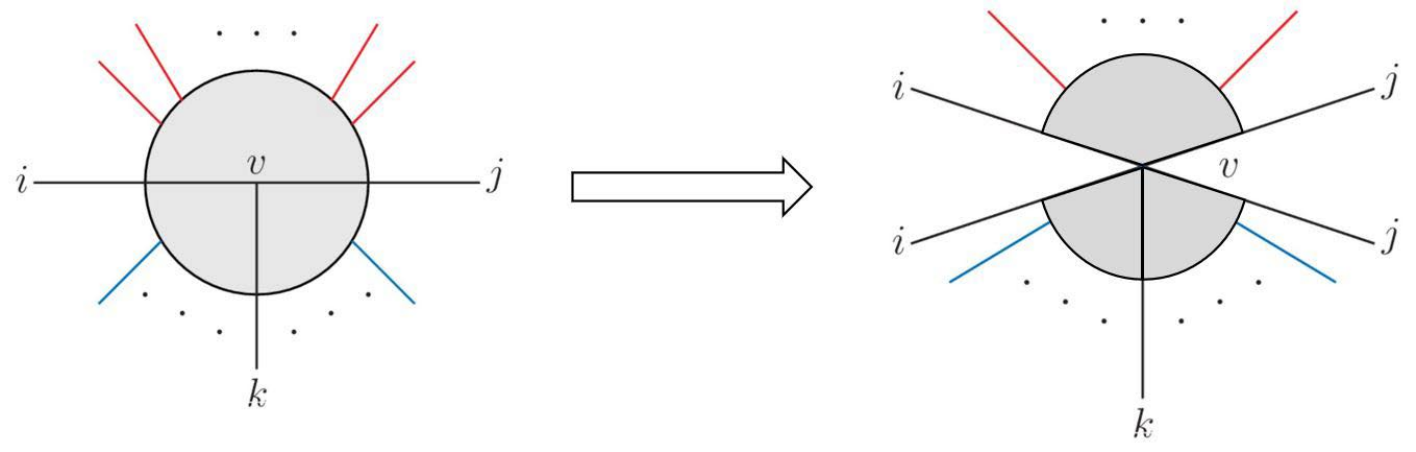} \\
  \caption{Diagrammatic interpretation for $2$-split: tearing $L_{(i,v)}$ and $L_{(j,v)}$.}\label{tear}
\end{figure}

From the perspective of Feynman diagrams, the $2$-split is complementary to the factorization on the poles used in the BCFW on-shell recursion relation. Both the BCFW method and the $2$-split require the selection of two special external legs $i$ and $j$. Let us assume that the choices of $i$ and $j$ in BCFW method and $2$-split are consistent, and focus on the line $L_{(i,j)}$ connecting $i$ and $j$. Factorization on a pole corresponds to cutting $L_{(i,j)}$ at an internal line; according to the interpretation in \cite{Zhou:2024ddy,Zhou:2026fkg}, the $2$-split can instead be understood as tearing $L_{(i,j)}$ apart along $L_{(i,v)}$ and $L_{(j,v)}$, much like opening two zippers, as illustrated in Fig.\ref{tear}. Here the vertex $v$ is the intersection point of $L_{(i,j)}$ with $L_{(k,v)}$, that is, for any connected Feynman diagram, one can always find a point $v$ on $L_{(i,j)}$ such that $L_{(i,v)}$, $L_{(j,v)}$ and $L_{(k,v)}$ meet at $v$. For a given Feynman diagram, all internal lines contained in $L_{(i,j)}$ provide all possible ways of cutting $L_{(i,j)}$, while all vertices on $L_{(i,j)}$ provide all possible ways of tearing $L_{(i,j)}$ apart, where the different choices of $v$ can be realized by the different choices of $k$. In this sense, one can regard the former as factorization on internal lines and the latter as factorization at vertices (edges of internal lines).
Thus, if we ask how many ways there are to achieve the factorization of the amplitude along the line connecting the external legs $i$ and $j$, then factorization on poles provides only a subset of them, while the $2$-split supplies the remainder. Owing to the above complementarity, it is reasonable to expect that the $2$-split provides the missing part of the original BCFW recursion relation. This observation is one of our motivations for choosing to determine the boundary term via the $2$-split.

%

%%%%%%%%%%%%%%%%%%%%%%%%%%%%%%
\subsection{From on-shell amplitudes to off-shell currents}
\label{subsec-onshell-current}
%%%%%%%%%%%%%%%%%%%%%%%%%%%%%%

At the end of the previous subsection, we mentioned that the complementary relationship between $2$-split and factorization at the poles motivates an attempt to determine the boundary term of the BCFW recursion relation from $2$-split. What we aim for is to develop an on-shell method, that is, a procedure in which only lower-point on-shell amplitudes are used throughout. However, the two currents generated by $2$-split are both off-shell, each carrying an off-shell external leg $\kappa$ or $\kappa'$. Therefore, in this subsection we will clarify how to obtain the expressions for the two currents from the expressions for the on-shell amplitudes.

In each current, there are three special external legs, $\{i,j,\kappa\}$ or $\{i,j,\kappa'\}$. Let us take the current ${\cal J}_{n_1}$ containing $\{i,j,\kappa\}$ as the example. Suppose that the expression for the amplitude when $k_\kappa$ is on-shell is ${\cal A}_{n_1}$. When $k_\kappa$ is off-shell, the procedure for obtaining the expression for the off-shell current ${\cal J}_{n_1}$ from the on-shell amplitude ${\cal A}_{n_1}$ is as follows \cite{Azevedo:2025vxo}. In the first step, all Mandelstam variables in ${\cal A}_{n_1}$ are decomposed into sums of two-point Mandelstam variables $s_{pq}=2k_p\cdot k_q$. In the second step, all $k_\kappa$ in $k_p\cdot k_\kappa$ among these two-point Mandelstam variables are replaced by the off-shell $k_\kappa$, namely,
\bea
s_{p\kappa}\,\to\,2k_p\cdot k_\kappa\,.~~~~\label{s-k}
\eea
Note that since $k_\kappa$ is off-shell, $2k_p\cdot k_\kappa$ can no longer be regarded as Mandelstam variables $s_{p\kappa}=(k_p+k_\kappa)^2$.
In the third step, the following shift is performed on $k_i\cdot k_j$, $k_i\cdot k_\kappa$ and $k_j\cdot k_\kappa$,
\bea
k_i\cdot k_j\,\to\,k_i\cdot k_j-{1\over 2}\,k_\kappa^2\,,~~~~k_i\cdot k_\kappa\,\to\,k_i\cdot k_\kappa+{1\over2}\,k_\kappa^2\,,~~~~
k_j\cdot k_\kappa\,\to\,k_j\cdot k_\kappa+{1\over2}\,k_\kappa^2\,.~~~~\label{shift-for-offshell-CHY}
\eea
After completing the above operations, one obtains the off-shell current ${\cal J}_{n_1}$. Carrying out the same operations on the off-shell external line $\kappa'$ as those performed on $\kappa$ yields the off-shell current ${\cal J}_{n_2}$.

The above shift \eref{shift-for-offshell-CHY} originates from the more general massive CHY formalism. When the external lines $i$, $j$ and $k$ are all massive, there exists a more general shift preserving the $SL(2,C)$ symmetry, thereby giving rise to the CHY formula for such massive tree amplitudes \cite{Naculich:2014naa,Naculich:2015zha}. In the case encountered here, both $i$ and $j$ are massless particles, so that the shift of the kinematic variables reduces to \eref{shift-for-offshell-CHY}. Since in \cite{Cao:2024qpp} the two currents ${\cal J}_{n_1}$ and ${\cal J}_{n_2}$ are defined through the CHY formula, the shift in \eref{s-k} and \eref{shift-for-offshell-CHY} is appropriate for both of these currents.

The above operations have the advantage that one can start from an arbitrary expression for the on-shell amplitude. Usually, when an on-shell momentum is replaced by an off-shell one, many relations satisfied by the on-shell Lorentz invariants are spoiled, so that expressions which were originally equivalent become inequivalent. However, the shift \eref{shift-for-offshell-CHY} restores the equivalence among different expressions for the on-shell amplitude. Let us perform a check using the $4$-point NLSM current ${\cal J}^{\rm NLSM}_4(i,p,j,\kappa)$ as an example. The on-shell amplitude ${\cal A}^{\rm NLSM}_4(i,p,j,\kappa)$ has the following equivalent expressions,
\bea
&&(1)~{\cal A}^{\rm NLSM}_4(i,p,j,\kappa)=s_{ip}+s_{pj}\,,~~~~~~(2)~{\cal A}^{\rm NLSM}_4(i,p,j,\kappa)=s_{pj}+s_{j\kappa}\,,\nn
&&(3)~{\cal A}^{\rm NLSM}_4(i,p,j,\kappa)=s_{j\kappa}+s_{\kappa i}\,,~~~~~~(4)~{\cal A}^{\rm NLSM}_4(i,p,j,\kappa)=s_{\kappa i}+s_{ip}\,,\nn
&&(5)~{\cal A}^{\rm NLSM}_4(i,p,j,\kappa)=-s_{ij}\,,~~~~~~~~~~~(6)~{\cal A}^{\rm NLSM}_4(i,p,j,\kappa)=-s_{p\kappa}\,.~~\label{6expressions-A}
\eea
By utilizing the shift in \eref{shift-for-offshell-CHY}, we get
\bea
&&(1)~{\cal J}^{\rm NLSM}_4(i,p,j,\kappa)=s_{ip}+s_{pj}\,,\nn
&&(2)~{\cal J}^{\rm NLSM}_4(i,p,j,\kappa)=s_{pj}+2k_j\cdot k_\kappa+k_\kappa^2=s_{pj}+(k_j+k_\kappa)^2=s_{pj}+(k_i+k_p)^2=s_{pj}+s_{ip}\,,\nn
&&(3)~{\cal J}^{\rm NLSM}_4(i,p,j,\kappa)=2k_j\cdot k_\kappa+2k_i\cdot k_\kappa+2k_\kappa^2=-2k_p\cdot k_\kappa=2k_p\cdot(k_i+k_p+k_j)=s_{ip}+s_{pj}\,,\nn
&&(4)~{\cal J}^{\rm NLSM}_4(i,p,j,\kappa)=2k_i\cdot k_\kappa+k_\kappa^2+s_{ip}=(k_i+k_\kappa)^2+s_{ip}=(k_j+k_p)^2+s_{ip}=s_{pj}+s_{ip}\,,\nn
&&(5)~{\cal J}^{\rm NLSM}_4(i,p,j,\kappa)=-s_{ij}+k_\kappa^2=2k_i\cdot(k_p+k_\kappa)+k_\kappa^2=s_{ip}+(k_i+k_\kappa)^2=s_{ip}+(k_j+k_p)^2=s_{ip}+s_{pj}\,,\nn
&&(6)~{\cal J}^{\rm NLSM}_4(i,p,j,\kappa)=-2k_p\cdot k_\kappa=2k_p\cdot(k_i+k_j)=s_{ip}+s_{pj}\,.~~\label{6expressions-J}
\eea
Thus we see that the six expressions for the current generated from the six expressions for the on-shell amplitude are all equivalent.

%%%%%%%%%%%%%%%%%%%%%%%%%%%%%%
\section{Determination of boundary terms}
\label{sec-fix-boundary}
%%%%%%%%%%%%%%%%%%%%%%%%%%%%%%

In this section, we develop a general method for computing the boundary terms of NLSM amplitudes recursively from lower-point on-shell amplitudes. We begin in subsection \ref{subsec-graph-boundary} by discussing the Feynman-diagrammatic structure corresponding to the boundary terms. Then, in subsection \ref{subsec-key-zero}, by combining the hidden zeros and the analysis of mass dimensions with the Feynman-diagrammatic structure obtained in subsection \ref{subsec-graph-boundary}, we derive the constraints that the boundary terms should satisfy. The analysis in this section will show that the NLSM$\oplus{\rm Tr}(\phi^3)$ amplitudes of interest in this paper have no boundary terms. Finally, in subsection \ref{subsec-key-split}, we investigate how to compute, via $2$-split, the boundary terms of the NLSM amplitudes that satisfy the constraints obtained in subsection \ref{subsec-key-zero}.

The discussion in subsection \ref{subsec-graph-boundary} applies to all tree amplitudes involving only scalar particles (including both real and virtual particles), while the discussion in subsection \ref{subsec-key-zero} concerns all ordered scalar amplitudes, with its final conclusions specialized to the NLSM$\oplus{\rm Tr}(\phi^3)$ and NLSM amplitudes of interest in this paper, and subsection \ref{subsec-key-split} focuses entirely on NLSM amplitudes.

%%%%%%%%%%%%%%%%%%%%%%%%%%%%%%%
\subsection{Diagrammatic feature of boundary term}
\label{subsec-graph-boundary}
%%%%%%%%%%%%%%%%%%%%%%%%%%%%%%%

At the end of section \ref{subsec-boundary}, we pointed out that the boundary terms of the BCFW recursion relation do not contain effective propagators along the line $L_{(i,j)}$.
In this subsection, we further analyze this feature and thereby determine the structure of Feynman diagrams corresponding to the boundary terms. The discussion in this subsection applies to all scalar amplitudes.

It is worth emphasizing that the Feynman diagrams discussed in this subsection are those whose associated Feynman rules correspond to the BCFW-shifted amplitude ${\cal A}_n(z)$, rather than the unshifted amplitude ${\cal A}_n$. The reason is that we need to extract the boundary term from ${\cal A}_n(z)$ in the manner described in section \ref{subsec-boundary}. That is, each $k_i$ or $k_j$ is replaced by $\hat{k}_i$ or $\hat{k}_j$ in \eref{BCFW-shift}.

We first note that the boundary term is defined as the contour integral at infinity of ${\cal A}_n(z)/z$. Therefore, if ${\cal A}_n(z)$ can be written as a sum of many terms, namely,
\bea
{\cal A}_n(z)\,=\,{\cal T}_1(z)+{\cal T}_2(z)+{\cal T}_2(z)+\cdots\,,~~~~\label{A-T}
\eea
then the boundary term satisfies
\bea
{\cal B}_n\,=\,{1\over2\pi i}\,\oint_{\infty}\,{dz\over z}\,{{\cal A}_n(z)}\,=\,{1\over2\pi i}\,\oint_{\infty}\,{dz\over z}\,{{\cal T}_1(z)}+{1\over2\pi i}\,\oint_{\infty}\,{dz\over z}\,{{\cal T}_2(z)}+{1\over2\pi i}\,\oint_{\infty}\,{dz\over z}\,{{\cal T}_3(z)}+\cdots\,.~~~~\label{A-T-boundary}
\eea
The above simple decomposition relation shows that the summation of several terms, for instance ${\cal T}_1(z)+{\cal T}_2(z)$, does not generate any new contributions to the boundary term of BCFW recursion relation beyond those already contained in ${\cal T}_1(z)$ and ${\cal T}_2(z)$.

Let us assume that each ${\cal T}_i(z)$ in \eref{A-T} arises from a specific Feynman diagram $\Gamma_i$. The decomposition relation \eref{A-T-boundary} then implies that the boundary term does not contain contributions that are absent in individual Feynman diagrams and arise only in the sum over Feynman diagrams. Thus, we can analyze the contribution of each individual Feynman diagram to the boundary term, with the complete boundary term being the sum of these contributions.

Now we consider the crucial feature that the boundary term contains no propagator along the line $L_{(i,j)}$. Based on the decomposition relation \eref{A-T-boundary}, we can study, within a single Feynman diagram, the contributions that contain no propagator along $L_{(i,j)}$, and finally sum the contributions of the different Feynman diagrams. It is easy to see that such contributions have two sources:
\begin{itemize}
\item (1) The Feynman diagram itself contains no internal line along $L_{(i,j)}$;
\item (2) Among the many terms produced by a single Feynman diagram, some terms have their numerators cancel the denominators, thereby removing the propagators along $L_{(i,j)}$.
\end{itemize}

\begin{figure}
  \centering
   \includegraphics[width=6cm]{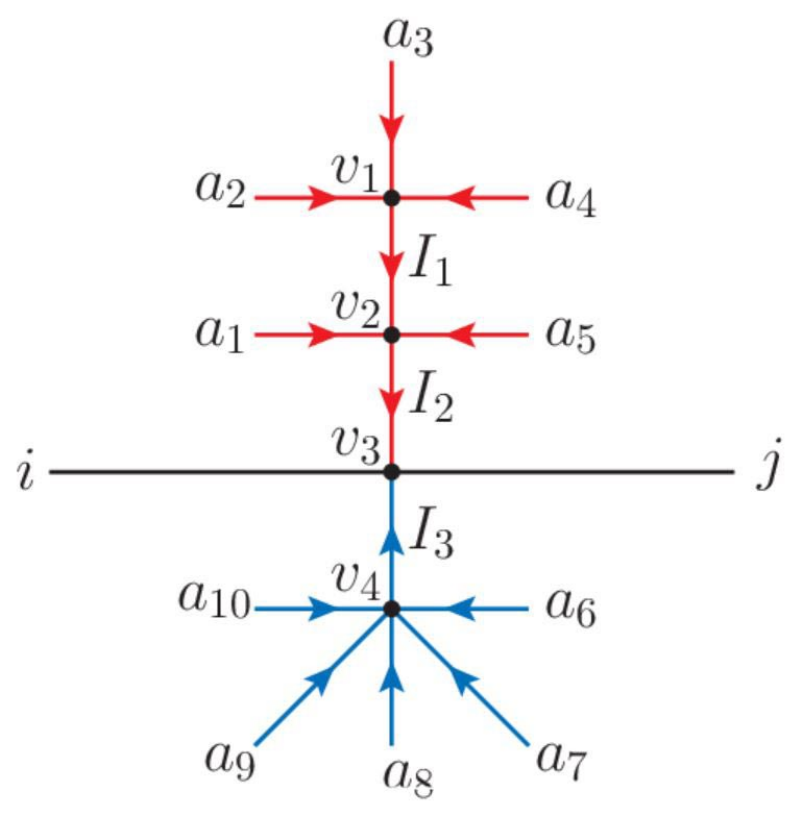} \\
  \caption{Direction of momenta flow. Level $1$: from external legs in $\{1,\cdots,n\}\setminus\{i,j\}$ to the line $L_{(i,j)}$.}\label{direction}
\end{figure}

We adopt the convention that all external momenta are incoming. To analyze the second case, in which the cancellations between numerators and denominators produce terms free of the propagator along $L_{(i,j)}$, we may, when writing down the terms corresponding to a Feynman diagram, remove one of $\hat{k}_i$ and $\hat{k}_j$ via momentum conservation. As will be seen shortly, this manipulation renders the cancellation process controllable. Without loss of generality, we assume that the external momentum eliminated is $\hat{k}_j$. This prescription means that, in any Feynman diagram, we need to write down the expressions for the vertices and propagators along the direction pointing from the remaining external legs to $j$. This directionality can be understood at two levels. At the first level, for the external lines in $\{1,\cdots,n\}\setminus\{i,j\}$, we write them along the direction pointing from these external legs to the line $L_{(i,j)}$. Taking Fig.\ref{direction} as an example, we need to write down the expressions for the vertices and propagators along the directions indicated by the arrows in the figure. For instance, for the vertex $v_1$, we understand the associated momenta as follows: the momenta $k_{a_2}$, $k_{a_3}$ and $k_{a_4}$ flow into $v_1$, while the momentum $k_{a_2}+k_{a_3}+k_{a_4}$ flows out of $v_2$. The same understanding applies to $v_2$ and $v_4$, with the momenta from the two sides of $L_{(i,j)}$ eventually flowing into the vertex $v_3$ on $L_{(i,j)}$. With this understanding, the propagators corresponding to the internal lines $I_1$, $I_2$ and $I_3$ are respectively
\bea
{1\over (k_{a_2}+k_{a_3}+k_{a_4})^2}\,,~~~~~~{1\over(k_{a_1}+k_{a_2}+k_{a_3}+k_{a_4}+k_{a_5})^2}\,,
~~~~~~{1\over(k_{a_6}+k_{a_7}+k_{a_8}+k_{a_9}+k_{a_{10}})^2}\,.
\eea
We now turn to consider the Lorentz invariants generated at each vertex. As shown in Fig.\ref{direction}, at each vertex there are several incoming momenta together with one outgoing momentum. Since $\hat{k}_j$ is removed, we need to use momentum conservation to eliminate the outgoing momentum and consider only the incoming momenta. Since we are considering scalar models, whereas the numerator of each scalar propagator is a constant and does not contain any $\eta_{\mu\nu}$, thus the momenta provided by each vertex can only contract into Lorentz invariants at that vertex itself. Thus, locality implies that the Lorentz invariants generated at the vertex $v_1$ depend only on $k_{a_2}$, $k_{a_3}$, $k_{a_4}$; the Lorentz invariants generated at the vertex $v_2$ depend only on $k_{a_1}$, $k_{a_5}$, $k_{I_1}=k_{a_2}+k_{a_3}+k_{a_4}$; and the Lorentz invariants generated at the vertex $v_4$ depend only on $k_{a_6}$, $k_{a_7}$, $k_{a_8}$, $k_{a_9}$, $k_{a_{10}}$. Note that in the Lorentz invariants generated at $v_2$, $k_{I_1}$ can only appear as a whole; its components, $k_{a_2}$, $k_{a_3}$ and $k_{a_4}$, cannot appear individually.
From the above discussion, we see that once $\hat{k}_j$ is required to be absent from the expression for the amplitude, then at the first level of directionality, namely in the process where momentum flows from the external legs in ${1,\cdots,n}\setminus\{i,j\}$ to $L_{(i,j)}$, no vertex generates a Lorentz invariant capable of cancelling any propagator on $L_{(i,j)}$.

\begin{figure}
  \centering
   \includegraphics[width=6cm]{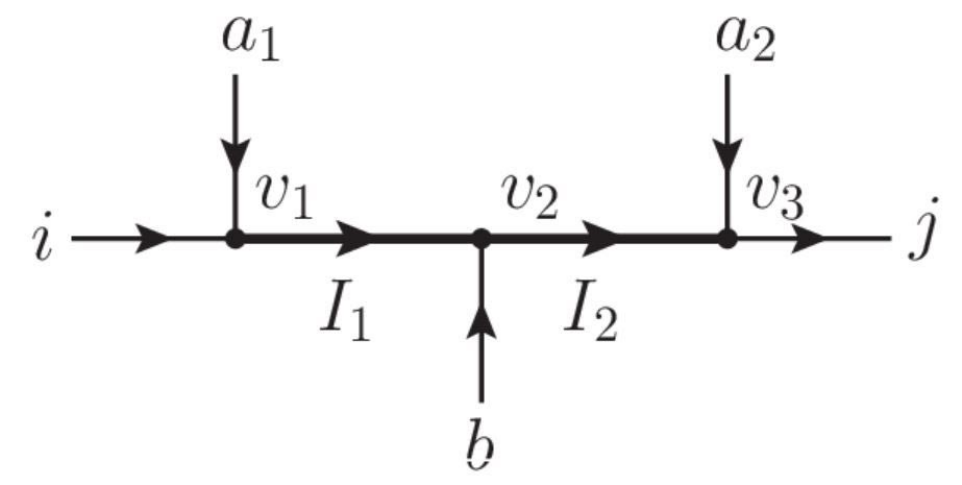} \\
  \caption{Direction of momenta flow. Level 2: from $i$ to $j$.}\label{direction(ij)}
\end{figure}

The second level of directionality is along the line $L_{(i,j)}$. Since it is assumed that the momentum $\hat{k}_j$ is removed through momentum conservation, we should describe the momentum flow along $L_{(i,j)}$ in the direction from $i$ to $j$, as shown in Fig.\ref{direction(ij)}.
Since we are considering a general scalar theory, we do not require all vertices to be of even valences as in the NLSM; thus trivalent vertices appear in Fig.\ref{direction(ij)}. As before, each vertex has several incoming momenta and one outgoing momentum. Since $\hat{k}_j$ is removed, we still eliminate the outgoing momentum via momentum conservation, so that the Lorentz invariant generated at each vertex depends only on the incoming momenta. We emphasize once again that, since we are considering scalar models, the numerator of each propagator contains no $\eta_{\mu\nu}$, the momenta provided by a given vertex can only be contracted at that vertex itself, rather than between different vertices, to form Lorentz invariants.
In this case, the Lorentz invariant generated by a vertex on $L_{(i,j)}$ may possibly cancel several propagators on $L_{(i,j)}$. For example, in Fig.\ref{direction(ij)}, the Lorentz invariant generated at vertex $v_2$ depends on $\hat{k}_{I_1}$ and $k_b$, and it is possible to contain $\hat{k}_{I_1}^2$ which cancels the propagator $1/\hat{k}_{I_1}^2$; similarly, the Lorentz invariant generated at vertex $v_3$ depends on $\hat{k}_{I_2}$ and $k_{a_2}$, and may contain $\hat{k}_{I_2}^2$ which cancels the propagator $1/\hat{k}_{I_2}^2$. In the above, $\hat{k}$ indicates that $\hat{k}_{I_1}$ and $\hat{k}_{I_2}$ depend on $z$.

\begin{figure}
  \centering
   \includegraphics[width=18cm]{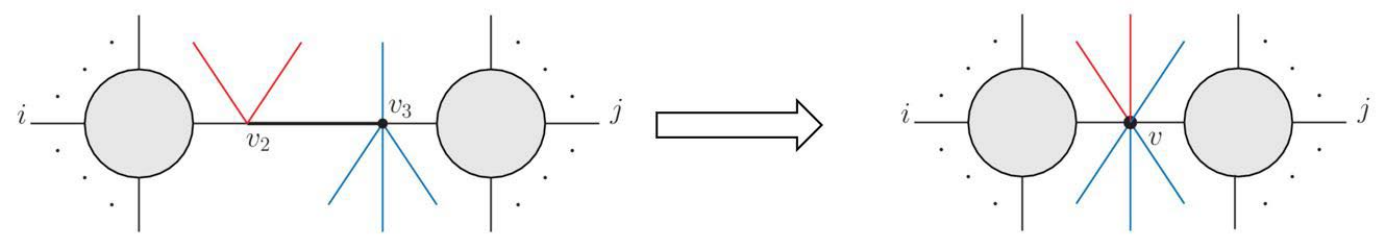} \\
  \caption{Shrinking the internal line connecting $v_1$ and $v_2$. The vertex $v_2$ carries $k_I^2$ which cancels the propagator between $v_1$ and $v_2$.}\label{shrink}
\end{figure}

We can split each vertex on $L_{(i,j)}$ into two parts, one capable of cancelling the adjacent propagator and the other not. Taking the vertex $v_3$ in Fig.\ref{direction(ij)} as an example, we split the expression for this vertex as
\bea
V_{3}\,=\,c_3\,\hat{k}_{I_2}^2+V^{\rm remain}_3\,,~~\label{split-v}
\eea
where the first part is proportional to $\hat{k}_{I_2}^2$, while the second part $V^{\rm remain}_3$ does not contain $\hat{k}_{I_2}^2$. From the diagrammatic point of view, this amounts to splitting one Feynman diagram into two, which have exactly the same topology but different expressions $c_3\,\hat{k}_{I_2}^2$ and $V^{\rm remain}_3$ at the vertex $v_3$.
In the first diagram, the $\hat{k}_{I_2}^2$ carried by the vertex
$v_3$ cancels the propagator connected to $v_3$, whereas the same vertex $v_3$ in the second diagram does not possess this property.
For the first diagram, the internal line between the two vertices $v_2$ and $v_3$ is effectively shrunk to a point, so that the two vertices merge into a new vertex $v$, as shown in Fig.\ref{shrink}. The expression for this $v$ is $V_2c_3$, where $V_2$ is the expression for the vertex $v_2$.
If $V_2$ is also decomposed into
\bea
V_2\,=\,c_2\,\hat{k}_{I_1}^2+V_2^{\rm remian}\,,
\eea
then the $c_2c_3\hat{k}_{I_1}^2$ part carried by $v$ can further cancel the propagator $1/\hat{k}_{I_1}^2$. Thus, the internal line $I_1$ is also shrunk, and $v_1$ and $v$ merge into a new single vertex $v'$ whose expression is $V_1c_2c_3$, where $V_1$ is the expression for the vertex $v_1$. In this case, after all internal lines have been shrunk, $L_{(i,j)}$ contains only a single vertex $v'$ and no propagators, which is characteristic of boundary terms.

In the above example, $c_3$ in \eref{split-v} may also contain $k_{I_2}^2$, namely,
\bea
c_3\,=\,c'_3\,\hat{k}_{I_2}^2+c_3^{\rm remain}\,.
\eea
The $c'_3\hat{k}_{I_2}^2$ part in $c_3$ can further participate in cancelling other propagators on $L_{(i,j)}$, for example $1/\hat{k}_{I_1}^2$ (since $\hat{k}_{I_2}^2=(\hat{k}_{I_1}+k_b)^2$ contains $\hat{k}_{I_1}^2$). When the $c'_3\hat{k}_{I_1}^2\hat{k}_{I_2}^2$ in $V_3$ cancels $1/\hat{k}_{I_1}^2$ and $1/\hat{k}_{I_2}^2$, then the three vertices $v_1$, $v_2$ and $v_3$ shrink into a single vertex $v$, and in this case the expression for $v$ is $V_1V_2c'_3$, where $V_1$ and $V_2$ are expressions for $v_1$ and $v_2$, respectively. This case also realizes the removal of all propagators on $L_{(i,j)}$, which is characteristic of boundary terms.

From the cancellations discussed above, we see that when $\hat{k}_j$ is removed via momentum conservation, the propagators on $L_{(i,j)}$ can only be canceled by the factors from the vertices on $L_{(i,j)}$. By extracting the factors at the vertices that participate in the cancellations, such processes can be understood from the diagrammatic perspective as shrinking the internal lines on $L_{(i,j)}$.

We now apply the above discussion of the cancellations to the second source of the boundary term: the cancellations between numerators and denominators render these terms free of propagators on $L_{(i,j)}$. According to the previous discussion, the diagrammatic structure for these terms can be viewed as shrinking the internal lines along $L_{(i,j)}$, and the resulting diagram has only a single vertex $v$ on $L_{(i,j)}$, thus matches the Feynman-diagrammatic structure corresponding to the first source. Since the cancellations occur only between the propagators and the vertices on $L_{(i,j)}$---a consequence of removing $\hat{k}_j$ from the expression---we arrive at the following conclusion. Assuming that there are $r$ internal lines $\{I_1,\cdots,I_r\}$ and $r+1$ vertices $\{v_1,\cdots,v_{r+1}\}$ on $L_{(i,j)}$. For a Feynman diagram obtained by shrinking internal lines and having only a single vertex $v$ on $L_{(i,j)}$, the mass dimension of $v$ is
\bea
[v]\,=\,-2r+\sum_{s=1}^{r+1}\,[v_s]\,.~~\label{dimension-v}
\eea
Vertices other than those on $L_{(i,j)}$ are completely unaffected by the cancellations. In general, the vertices $v$
can have different mass dimensions $[v]$ in different Feynman diagrams. However, as will be seen in subsequent subsections, for NLSM$\oplus{\rm Tr}(\phi^3)$ amplitudes with three ${\rm Tr}(\phi^3)$ external legs and the pure NLSM amplitudes, all Feynman diagrams corresponding to boundary terms have vertices $v$ of a common mass dimension.

In summary, all boundary terms arise from Feynman diagrams that have only a single vertex $v$ on $L_{(i,j)}$ and no internal line. Since some of these diagrams originate from shrinking internal lines and merging the nearby vertices, the Feynman rules for $v$ are in general rather messy and complicated, as exemplified $V_1c_2c_3$ and $V_1V_2c'_3$ from Fig.\ref{direction(ij)}. However, as will be seen later, we actually need only this diagrammatic structure and the mass dimension of $v$ found in \eref{dimension-v}, rather than the explicit Feynman rules.

%%%%%%%%%%%%%%%%%%%%%%%%%%%%%%%
\subsection{Constraints on boundary terms from hidden zeros and mass dimensions}
\label{subsec-key-zero}
%%%%%%%%%%%%%%%%%%%%%%%%%%%%%%%

In this subsection, we combine the hidden zeros and mass dimensions with the Feynman-diagrammatic structure obtained in the previous subsection to analyze the properties of the boundary terms. Most of the discussion in this section, as well as the resulting expansion of the boundary term in \eref{expan-doundary}, apply to arbitrary ordered scalar amplitudes. In the final part of this subsection, the analysis of $C_{ab}$ in \eref{expan-doundary} is specialized to the NLSM$\oplus{\rm Tr}(\phi^3)$ and NLSM amplitudes.

In subsection \ref{subsec-0-and-split}, we introduced the hidden zeros of NLSM amplitudes and of NLSM$\oplus{\rm Tr}(\phi^3)$ amplitudes involving three ${\rm Tr}(\phi^3)$ scalars. We now observe that when the kinematic condition \eref{kine-condi-0-scalar} for the hidden zero is satisfied, the boundary terms of these amplitudes also vanish. The reason is simple: the BCFW-shift in \eref{BCFW-shift} does not affect the zero condition \eref{kine-condi-0-scalar}, and therefore, when the condition \eref{kine-condi-0-scalar} is satisfied, ${\cal A}_n(z)$ likewise vanishes. If ${\cal A}_n(z)$ is expanded in $z$, each order in $z$ must vanish independently, and as shown in \eref{expan-A(z)}, the boundary term is precisely the $z^0$-order term, so it must vanish.

Based on the above observation, we now consider how the zero condition $s_{ab}=0$ in \eref{kine-condi-0-scalar} gives rise to the vanishing of the boundary term. To this end, we first express the amplitude ${\cal A}_n(z)$ in terms of independent Mandelstam variables, so as to avoid adding or removing $s_{ab}$ through momentum conservation and the on-shell conditions. In addition, we require that the chosen independent Mandelstam variables preserve all $s_{ab}$. Our choice is as follows. First, we choose only $2$-point Mandelstam variables $s_{pq}$. Then, in accordance with the conventions adopted in the previous subsection, we eliminate $\hat{k}_j$ by momentum conservation.
After removing $\hat{k}_j$, the number of remaining $2$-point Mandelstam variables is $(n-1)(n-2)/2$. However, the number of independent Mandelstam variables is
\bea
{n(n-3)\over2}\,=\,{(n-1)(n-2)\over2}-1\,,
\eea
because the remaining $2$-point Mandelstam variables satisfy the constraint
\bea
\sum_{p',q'\in\{1,\cdots,n\}\setminus j}\,s_{p'q'}=\hat{k}_j^2=0\,.
\eea
Therefore, in order to retain as much of $s_{ab}$ as possible while choosing independent Mandelstam variables, we further remove one $2$-point Mandelstam variable $\hat{s}_{it}$, where $t\in\{1,\cdots,n\}\setminus\{i,j\}$. As in the case of $\hat{k}$, here $\hat{s}$ denotes that this Mandelstam variable depends on $z$.
Now, the remaining $n(n-3)/2$ Mandelstam variables are all independent, and all $s_{ab}$ are contained therein.
In summary, the set of chosen Mandelstam variables is
\bea
\{s_{pq}\}\setminus\{\{\hat{s}_{pj}\}\cup\hat{s}_{it}\}\,.~~~~\label{M-variable}
\eea

When ${\cal A}_n(z)$ is expressed in terms of the Mandelstam variables in \eref{M-variable}, we can make the following important observation.
Since the boundary term vanish under the kinematic condition
\eref{kine-condi-0-scalar}, it follows that for any ordered scalar amplitude the boundary term must be expressible as
\bea
{\cal B}_n\,=\,\sum_{a\in A,b\in B}\,s_{ab}\,C_{ab}\,.~~~~\label{expan-doundary}
\eea
The argument is simple. We may first single out a particular
pair $(a_0,b_0)$ where $a_0\in A$, $b_0\in B$, and then collect the components in the boundary term that contain the remaining
$s_{ab}$ belonging to $\{s_{ab}\}\setminus s_{a_0b_0}$. That is, we write the boundary term as
\bea
{\cal B}_n\,=\,D+\sum_{\{s_{ab}\}\setminus s_{a_0 b_0}}\,s_{ab}\,C_{ab}\,.
\eea
Clearly, if $D$ itself does not vanish, it must contain
$s_{a_0b_0}$. The reason is that the entire boundary term vanishes under the kinematic condition $s_{ab}=0$, so
$D$ must also vanish under this condition, since the second part on the r.h.s. vanishes automatically. This demonstrates that the boundary term can be expanded as
\eref{expan-doundary}. Note that each $C_{ab}$ is allowed to be zero. That is, we only assume that the amplitude vanishes under the zero condition \eref{kine-condi-0-scalar}, without making any prior requirement on how much $s_{ab}$ the boundary term contains.
In an ordered amplitude, the particles contained in the sets
$A$ and $B$ are separated by $i$ and $j$, thus no propagator of the form
$1/s_{ab}$ can arise. Therefore we need not worry about the possibility that the
$s_{ab}$ in the numerator is cancelled by the denominator, nor about the divergence induced by
$1/s_{ab}$ on the locus $s_{ab}=0$ in kinematic space.

Combining the Feynman-diagrammatic structure of the boundary term obtained in the previous subsection \ref{subsec-graph-boundary} with dimensional analysis, one can derive certain constraints that
$C_{ab}$ in \eref{expan-doundary} must satisfy. In choosing the independent
Mandelstam variables, we have adopted the convention that
$\hat{k}_j$ is removed by momentum conservation, in complete agreement with the treatment in the previous subsection. Therefore, the analysis of the momentum-flow directions and of the Feynman-diagrammatic structure of the boundary term in the previous subsection applies equally well to our choice of independent Mandelstam variables in \eref{M-variable}. Since the momenta flow from the external legs in
$\{1,\cdots,n\}\setminus j$ towards $j$, according to the discussion in the previous subsection,
$s_{ab}$ can only be generated at the vertices on $L_{(i,j)}$. Vertices not on $L_{(i,j)}$ can only produce $s_{a_1a_2}$ or $s_{b_1b_2}$, where $a_1,a_2\in A$, $b_1,b_2\in B$.
For each Feynman diagram corresponding to the boundary term, there is no internal line on $L_{(i,j)}$, but only a single vertex
$v$; hence all the
$s_{ab}$ in the boundary term originate from this vertex $v$. Consequently, the mass dimension of the vertex
$v$ imposes a strong constraint on
$C_{ab}$ in \eref{expan-doundary}. For instance, if the mass dimension of the vertex
$v$ is $0$, no $s_{ab}$ can be generated at $v$, therefore all $C_{ab}$ must vanish, which means that no boundary term exists.

We now analyze, along the above line, the boundary terms of the
NLSM$\oplus{\rm Tr}(\phi^3)$ amplitudes involving three
${\rm Tr}(\phi^3)$ scalar particles, and of the NLSM amplitudes. For the former, we assume that both external legs
$i_\phi$ and $j_\phi$ are ${\rm Tr}(\phi^3)$ scalars. The vertices of
NLSM$\oplus{\rm Tr}(\phi^3)$ model have two possible mass dimensions \cite{Low:2017mlh,Yin:2018hht,Low:2018acv}. If three ${\rm Tr}(\phi^3)$ scalar lines are attached to a vertex, its mass dimension is $0$; otherwise, its mass dimension is $2$ (a vertex can be attached to at most three ${\rm Tr}(\phi^3)$ scalar lines). According to the discussion in the previous subsection, diagrams with only a single vertex
$v$ on the line $L_{(i,j)}$ arise from two sources. The first is that the Feynman diagrams already possess this structure from the outset. In this case, three ${\rm Tr}(\phi^3)$ scalar lines and
several pion lines are attached to $v$. Such a vertex has mass dimension $0$. The second source is that the
kinematic factors provided by vertices on
$L_{(i,j)}$ cancel the propagators on $L_{(i,j)}$. If there are $r$ internal lines on $L_{(i,j)}$, then there must be $r+1$ vertices. In order to cancel all the propagators on $L_{(i,j)}$, at least $r$ of the $r+1$ vertices must have mass dimension $2$. On the other hand, since $i_\phi$ and $j_\phi$ are ${\rm Tr}(\phi^3)$ scalars, there must exist on $L_{(i,j)}$ a vertex connecting three ${\rm Tr}(\phi^3)$ scalar lines. Therefore, there are $r$ vertices of mass dimension $2$ and one vertex of mass dimension $0$ on $L_{(i,j)}$. Using the observation \eref{dimension-v}, we see that the mass dimension of the vertex $v$ is $0$. This shows that, for diagrams of the second type, the mass dimension of the vertex $v$ on $L_{(i,j)}$ is still $0$. Therefore, for the NLSM$\oplus{\rm Tr}(\phi^3)$ amplitudes under consideration, no boundary term exists. In section \ref{sec-example-nlsm+phi}, based on this conclusion, we will compute two $5$-point
NLSM$\oplus{\rm Tr}(\phi^3)$ amplitudes using the BCFW recursion relation, and compare the results with those obtained from Feynman rules, thereby verifying the assertion that the boundary terms vanish.

Next we turn to $C_{ab}$ in the boundary terms
of the NLSM amplitudes. In the NLSM, all vertices have mass dimension
$2$. The observation \eref{dimension-v} then implies that the single vertex $v$ on
$L_{(i,j)}$ has mass dimension $2$, irrespective of whether the diagram is an original diagram or one obtained through cancellations between numerators and denominators. Therefore, any term generated at $v$ contains at most one $s_{ab}$. Since no other vertex can generate
$s_{ab}$, it follows that when we expand the boundary term of the NLSM amplitude in the form
\eref{expan-doundary}, each $C_{ab}$ depends only on $s_{a_1a_2}$ and $s_{b_1b_2}$, where $a_1,a_2\in A$, $b_1,b_2\in B$. That is to say, $C_{ab}$ is independent of any $s_{a'b'}$, where $a'\in A$, $b'\in B$. In the next subsection, based on this observation, we will construct $C_{ab}$ of the NLSM amplitude by utilizing the $2$-split behavior.

%%%%%%%%%%%%%%%%%%%%%%%%%%%%%%%
\subsection{From $2$-split to recursion relation of NLSM boundary terms}
\label{subsec-key-split}
%%%%%%%%%%%%%%%%%%%%%%%%%%%%%%%

We have argued that when a scalar amplitude satisfies the hidden zero described by
\eref{zero-scalar}, its boundary term also vanishes under the same kinematic condition. However, if we remove some
$s_{a''b''}$ from the zero condition \eref{kine-condi-0-scalar}, i.e. let $s_{a''b''}\neq0$ where $a''\in A$, $b''\in B$, then nonzero terms may appear in the expansion \eref{expan-doundary} of the boundary term. If we have a way to compute the nonzero part of the boundary term
${\cal B}_n$ in the above case, this amounts to probing a subset $\{C_{a''b''}\}\subset\{C_{ab}\}$. For the NLSM amplitude, since
each $C_{ab}$ contains no $s_{a'b'}$, all $C_{a''b''}$ can be probed in this way, and there is no need to worry about overlaps between the probed terms.

The novel factorization near the hidden zeros, namely the behavior termed $2$-split introduced in section \ref{subsec-0-and-split}, provides a route to realizing the above idea. The $2$-split condition \eref{kine-condi-split-scalar} is achieved by choosing an external leg $k$ from the set
$B$ and removing all $s_{ak}$ from the zero condition \eref{kine-condi-0-scalar}, these $s_{ak}$ correspond to the $s_{a''b''}$ in the above idea. To push this idea further, we denote by
$B^k_n$ the boundary term subject to the $2$-split condition \eref{kine-condi-split-scalar}, i.e.,
\bea
{\cal B}^{\rm NLSM}_n\,&\xrightarrow[]{\eref{kine-condi-split-scalar}}&\,B^k_n\,.~~\label{B-Bk}
\eea
Clearly, $B^k_n$ can be expanded in a form analogous to \eref{expan-doundary}, that is,
\bea
B^k_n\,=\,\sum_{a\in A}\,s_{ak}\,C_{ak}\,.~~\label{expan-Bk}
\eea
Although $B^k_n$ is subject to the $2$-split condition \eref{kine-condi-split-scalar}, since
$C_{ab}$ contains no $s_{a'b'}$, the $2$-split condition has no effect on
$C_{ak}$. Therefore, each $C_{ak}$ in \eref{expan-Bk} is identical to the corresponding $C_{ak}$ in \eref{expan-doundary}.
On the other hand, if we denote by $A^k_n(z)$ the amplitude ${\cal A}_n(z)$ subject to the $2$-split condition
\eref{kine-condi-split-scalar}, i.e.,
\bea
{\cal A}_n(z)\,&\xrightarrow[]{\eref{kine-condi-split-scalar}}&\,A^k_n(z)\,,~~\label{A-Ak}
\eea
then $B^k_n$ is also the boundary term of $A^k_n$, satisfying
\bea
B^k_n\,=\,{1\over 2\pi i}\,\oint_\infty\,{dz\over z}\,A^k_n(z)\,.~~~~\label{Ak-Bk}
\eea
This is because
the contour integration around the infinity commutes with imposing the $2$-split condition \eref{kine-condi-split-scalar}.
Therefore, we can probe all $C_{ak}$ by computing the boundary term of $A^k_n$. If
$k$ is taken to run over all external legs in the set
$B$, then all $C_{ab}$ can be probed, thus the complete boundary term can be obtained as
\bea
{\cal B}^{\rm NLSM}_n\,=\,\sum_{k\in B}\,B^k_n\,=\,{1\over2\pi i}\,\sum_{k\in B}\,\oint_\infty\,{dz\over z}\,A^k_n(z)\,,~~~~\label{sumk-forB}
\eea
due to the observation
\bea
\sum_{k\in B}\,B^k_n\,=\,\sum_{k\in B}\,\Big(\sum_{a\in A}\,s_{ak}\,C_{ak}\Big)\,=\,\sum_{a\in A,b\in B}\,s_{ab}\,C_{ab}\,=\,{\cal B}^{\rm NLSM}_n\,.
\eea

Through the $2$-split behavior of the NLSM amplitude in \eref{2split-scalar}, we can express
$A^k_n$ as the product of two currents ${\cal J}_{n_1}$ and ${\cal J}_{n_2}$ (as can be seen by plugging \eref{2split-scalar} into \eref{A-Ak}). Moreover, through the shift in \eref{s-k} and \eref{shift-for-offshell-CHY}, the off-shell currents ${\cal J}_{n_1}$ and ${\cal J}_{n_2}$ can be obtained from the two on-shell amplitudes
${\cal A}_{n_1}$ and ${\cal A}_{n+3-n_1}$, namely,
\bea
A^k_n(z)\,=\,{\cal A}^{\kappa-{\rm shift}}_{n_1}(\hat{i},\cdots,\hat{j},\kappa)\,\times\,{\cal A}^{\kappa'-{\rm shift}}_{n+3-n_1}(\hat{i},\hat{j},B(\kappa'))\,.~~~~\label{Ak-expression}
\eea
In the above, the superscript $\kappa$-shift denotes performing the shift \eref{shift-for-offshell-CHY} on
$\{\hat{i},\hat{j},\kappa\}$ in ${\cal A}_{n_1}(z)$, while $\kappa'$-shift denotes performing this shift on
$\{\hat{i},\hat{j},\kappa'\}$ in ${\cal A}_{n+3-n_1}(z)$. The notations $\hat{i}$ and $\hat{j}$ mean $k_i$ and $k_j$ are deformed as in \eref{BCFW-shift}. It is easy to see that the BCFW-shift in \eref{BCFW-shift}, and the shift of Mandelstam variables in \eref{s-k} and \eref{shift-for-offshell-CHY}, do not affect each other.
As explained around \eref{2split-scalar}, when
$n_1$ is even, the external legs $i_\phi,j_\phi,\kappa'_\phi$ in ${\cal A}_{n+3-n_1}(z)$ are interpreted as
${\rm Tr}(\phi^3)$ scalars, whereas when $n_1$ is odd, the external legs $i_\phi,j_\phi,\kappa_\phi$ in
${\cal A}_{n_1}(z)$ are understood as ${\rm Tr}(\phi^3)$ scalars. All remaining external particles are pions.
Since an ordered amplitude is endowed with a natural choice of the sets $A$ and $B$, once the pair of external legs
$(i,j)$ is chosen, $n_1$ is thereby fixed, so no sum over $n_1$ is needed in \eref{Ak-expression}.
Finally, we note that since the $2$-split holds only on the special locus \eref{kine-condi-split-scalar} in kinematic space, when we express $k_\kappa$ and $k_{\kappa'}$ in terms of the external momenta of the full amplitude, these external momenta must also satisfy the $2$-split condition
\eref{kine-condi-split-scalar}.

By plugging \eref{Ak-expression} into \eref{sumk-forB}, we finally obtain a simple formula for computing the boundary term of an $n$-point NLSM amplitude
from lower-point on-shell amplitudes
\bea
{\cal B}_n^{\rm NLSM}\,=\,{1\over 2\pi i}\,\sum_{k\in B}\,\oint_\infty\,{dz\over z}\,\Big({\cal A}^{\kappa-{\rm shift}}_{n_1}(\hat{i},\cdots,\hat{j},\kappa)\,\times\,{\cal A}^{\kappa'-{\rm shift}}_{n+3-n_1}(\hat{i},\hat{j},B(\kappa'))\Big)\Big|^{\eref{kine-condi-split-scalar}}_{\eref{M-variable}}.~~\label{recursionform-boundary}
\eea
The superscript \eref{kine-condi-split-scalar} emphasizes that the kinematic variables in ${\cal A}^{\kappa-{\rm shift}}_{n_1}(\hat{i},\cdots,\hat{j},\kappa)$ and ${\cal A}^{\kappa'-{\rm shift}}_{n+3-n_1}(\hat{i},\hat{j},B(\kappa'))$ should satisfy the $2$-split condition \eref{kine-condi-split-scalar}. The subscript
\eref{M-variable} indicates that the integrands should be expressed in terms of the Mandelstam variables chosen in \eref{M-variable}. This is because the analysis of the properties of the boundary term in the previous subsections \ref{subsec-graph-boundary} and \ref{subsec-key-zero} was based on this choice of the Mandelstam variables. If one replace the Mandelstam variables in ${\cal B}_n^{\rm NLSM}$ and ${\cal A}^{\kappa-{\rm shift}}_{n_1}(z)\times{\cal A}^{\kappa'-{\rm shift}}_{n+3-n_1}(z)$ by using momentum conservation and the on-shell conditions, the transformation undergone by ${\cal B}_n^{\rm NLSM}$ is generic; however, since the $2$-split in \eref{2split-scalar} holds only under the kinematic constraint \eref{kine-condi-split-scalar}, if the factorization behavior \eref{Ak-expression} is to remain valid, the transformation applied to ${\cal A}^{\kappa-{\rm shift}}_{n_1}(z)\times{\cal A}^{\kappa'-{\rm shift}}_{n+3-n_1}(z)$ must also be performed under the constraint \eref{kine-condi-split-scalar}. In other words, the transformations experienced by ${\cal B}_n^{\rm NLSM}$ and ${\cal A}^{\kappa-{\rm shift}}_{n_1}(z)\times{\cal A}^{\kappa'-{\rm shift}}_{n+3-n_1}(z)$ are not equivalent, and this will spoil many equivalence relations.

A straightforward way to achieve the chosen Mandelstam variables is to express ${\cal A}^{\kappa-{\rm shift}}_{n_1}$ and ${\cal A}^{\kappa'-{\rm shift}}_{n+3-n_1}$ in terms of the Mandelstam variables
\bea
\eref{M-variable}\,\xrightarrow[]{\hat{s}\to s}\,\{s_{pq}\}\setminus\{\{s_{pj}\}\cup s_{it}\}\,,~~~~\label{M-noz}
\eea
where ${\cal A}^{\kappa-{\rm shift}}_{n_1}$ and ${\cal A}^{\kappa'-{\rm shift}}_{n+3-n_1}$ are the currents before the BCFW-shift and do not depend on $z$.
Alternatively, one can start the calculation from arbitrary initial expressions for ${\cal A}^{\kappa-{\rm shift}}_{n_1}(z)$ and ${\cal A}^{\kappa'-{\rm shift}}_{n+3-n_1}(z)$. The resulting expression, however, must be transformed by imposing the $2$-split condition \eref{kine-condi-split-scalar} and made consistent with the Mandelstam variables in \eref{M-variable}, thereby recovering the expansion formula of the boundary term in \eref{expan-doundary}. The qualified ${\cal B}_n$ is not entirely expressed in terms of the Mandelstam variables in \eref{M-variable}, because $s_{ip}$ and $k_p\cdot q$ in $\hat{s}_{ip}$ have different powers of
$z$ and hence contribute differently to the boundary term. However, since $\hat{s}_{pj}$ and $\hat{s}_{it}$ are removed,
${\cal B}_n$ cannot contain $s_{pj}$, $s_{it}$, and $k_t\cdot q$. Note that removing $\hat{s}_{pj}$ does not remove the corresponding $k_p\cdot q$, since $-\hat{s}_{ip}$ also provides this term. Therefore, what the second approach amounts to is to bring
${\cal B}_n$ into the form described above, under the constraint of the $2$-split condition \eref{kine-condi-split-scalar}.

Whichever path is employed, one must first compute, under the $2$-split condition \eref{kine-condi-split-scalar}, the boundary term expressed in terms of appropriate kinematic variables, before the boundary term can be allowed to change its expression free from the constraint from $2$-split condition \eref{kine-condi-split-scalar}. In section \ref{sec-example-nlsm}, we will further demonstrate this point through examples.

In practical calculations, we do not need to evaluate the contour integral in \eref{recursionform-boundary}. A more efficient approach is, as in \eref{expan-A(z)}, to expand ${\cal A}^{\kappa-{\rm shift}}_{n_1}(z)\times{\cal A}^{\kappa'-{\rm shift}}_{n+3-n_1}(z)$ in $z$ and extract the constant term. It should be emphasized that we do not compute the boundary terms of ${\cal A}^{\kappa-{\rm shift}}_{n_1}$ and ${\cal A}^{\kappa'-{\rm shift}}_{n+3-n_1}$ separately and multiply them; rather, we compute the boundary term of the full ${\cal A}^{\kappa-{\rm shift}}_{n_1}\times{\cal A}^{\kappa'-{\rm shift}}_{n+3-n_1}$, then sum over all $k$. In section \ref{sec-example-nlsm}, we will use the recursive formula \eref{recursionform-boundary} to compute the boundary terms of $6$-point NLSM amplitudes.

%%%%%%%%%%%%%%%%%%%%%%%%%%%%%%
\section{Examples: NLSM$\oplus{\rm Tr}(\phi^3)$ amplitudes}
\label{sec-example-nlsm+phi}
%%%%%%%%%%%%%%%%%%%%%%%%%%%%%%

In this section, we compute two $5$-point NLSM$\oplus{\rm Tr}(\phi^3)$ amplitudes ${\cal A}_5^{{\rm NLSM}\oplus{\rm Tr}(\phi^3)}(1_\phi,2_\phi,3_\phi,4_\pi,5_\pi)$ and ${\cal A}_5^{{\rm NLSM}\oplus{\rm Tr}(\phi^3)}(1_\phi,2_\phi,3_\pi,4_\phi,5_\pi)$ using the BCFW on-shell recursion relation. There are two reasons for computing these two examples. First, in the next section, we will use the recursive formula \eref{recursionform-boundary} to compute the boundary terms of the $6$-point NLSM amplitudes, and the two NLSM$\oplus{\rm Tr}(\phi^3)$ amplitudes to be computed in this section will appear as lower-point amplitudes in \eref{recursionform-boundary}. Since our goal is to compute higher-point on-shell amplitudes from lower-point on-shell amplitudes throughout, it is unsatisfactory to leave the lower-point amplitudes required in \eref{recursionform-boundary} to be computed by Feynman rules; we would like them to be computable via the on-shell recursive method as well. The second reason is that in section \ref{subsec-key-zero} we found that for each NLSM$\oplus{\rm Tr}(\phi^3)$ amplitude with three ${\rm Tr}(\phi^3)$ scalar external legs, if we chose $i$ and $j$ as ${\rm Tr}(\phi^3)$ scalars, then the boundary term vanishes. We wish to verify this conclusion in concrete examples, that is, based on this conclusion, to compute the $5$-point NLSM$\oplus{\rm Tr}(\phi^3)$ amplitudes via the BCFW method and compare them with the results obtained from Feynman rules. As will be seen, the results obtained in this section are in complete agreement with those given by Feynman rules, and hence the conclusion obtained in section \ref{subsec-key-zero} is correct.

%%%%%%%%%%%%%%%%%%%%%%%%%%%%%%
\subsection{$5$-point amplitude ${\cal A}_5^{{\rm NLSM}\oplus{\rm Tr}(\phi^3)}(1_\phi,2_\phi,3_\phi,4_\pi,5_\pi)$}
\label{subsec-nlsm+phi-example1}
%%%%%%%%%%%%%%%%%%%%%%%%%%%%%%

The first example is the $5$-point amplitude ${\cal A}_5^{{\rm NLSM}\oplus{\rm Tr}(\phi^3)}(1_\phi,2_\phi,3_\phi,4_\pi,5_\pi)$, which carries
three ${\rm Tr}(\phi^3)$ scalars $1_\phi$, $2_\phi$, $3_\phi$, and two pions $4_\pi$, $5_\pi$. To apply the BCFW method, we choose $i=1$, $j=3$,
thus the BCFW-shift in \eref{BCFW-shift} becomes
\bea
\hat{k}_1=k_1+zq\,,~~~~~~\hat{k}_3=k_3-zq\,,~~~~~~~~{\rm where}~k_1\cdot q=k_3\cdot q=q^2=0\,.
\eea
Then the recursive formula \eref{BCFW-recursive-form} yields
\bea
{\cal A}_5^{{\rm NLSM}\oplus{\rm Tr}(\phi^3)}(1_\phi,2_\phi,3_\phi,4_\pi,5_\pi)\,=&&{{\cal A}^{{\rm Tr}(\phi^3)}_3(\hat{1},2,\hat{I})\,
{\cal A}^{\rm NLSM}_4(\hat{I},\hat{3},4,5)\Big|_{\hat{s}_{12}(z^*_1)=0}\over s_{12}}\nn
&+&{{\cal A}^{\rm NLSM}_4(4,5,\hat{1},\hat{I})\,
{\cal A}^{{\rm Tr}(\phi^3)}_3(\hat{I},2,\hat{3})\Big|_{\hat{s}_{451}(z^*_2)=0}\over s_{451}}\nn
&+&{\cal B}_5^{{\rm NLSM}\oplus{\rm Tr}(\phi^3)}(\hat{1}_\phi,2_\phi,\hat{3}_\phi,4_\pi,5_\pi)\,.~~~\label{nlsm+phi-1-recur}
\eea
According to the conclusion obtained in section \ref{subsec-key-zero}, the boundary term ${\cal B}_5^{{\rm NLSM}\oplus{\rm Tr}(\phi^3)}$ in the last line vanishes.
Therefore, the full amplitude can be computed from the recursive parts given by the first and second lines.

In the first and second lines, $z_1^*$ and $z_2^*$ can be solved as
\bea
\hat{s}_{12}=0~~&\to&~~z_1^*={-s_{12}\over 2k_2\cdot q}\,,\nn
\hat{s}_{451}=0~~&\to&~~z_2^*={-s_{451}\over 2K_{45}\cdot q}\,,~~~~\label{solve-z-example1}
\eea
where we have introduced the notation
\bea
K_{\a}=\sum_{p\in\a}\,k_p\,,~~\label{defin-K}
\eea
for any set $\a$ of external legs.
The on-shell amplitudes at lower points take the following forms,
\bea
&&{\cal A}^{{\rm Tr}(\phi^3)}_3(\hat{1},2,\hat{I})\,=\,{\cal A}^{{\rm Tr}(\phi^3)}_3(\hat{I},2,\hat{3})\,=\,1\,,\nn
&&{\cal A}^{\rm NLSM}_4(\hat{I},\hat{3},4,5)\,=\,-\hat{s}_{35}\,,~~~~~~~~{\cal A}^{\rm NLSM}_4(4,5,\hat{1},\hat{I})\,=\,-\hat{s}_{41}\,.~~\label{loweramp-example1}
\eea
Substituting the solutions in \eref{solve-z-example1} into \eref{loweramp-example1}, we arrive at
\bea
{\cal A}^{\rm NLSM}_4(\hat{I},\hat{3},4,5)\,&=&\,-s_{35}-s_{12}\,{k_5\cdot q\over k_2\cdot q}\,,\nn
{\cal A}^{\rm NLSM}_4(4,5,\hat{1},\hat{I})\,&=&\,-s_{14}+s_{451}\,{k_4\cdot q\over K_{45}\cdot q}\,.~~\label{lower-substiz-example1}
\eea
By plugging \eref{loweramp-example1} and \eref{lower-substiz-example1} into \eref{nlsm+phi-1-recur}, and using the observation
\bea
q\cdot k_2\,=\,-q\cdot K_{1345}\,=\,-q\cdot K_{45}\,,
\eea
we get the final result
\bea
{\cal A}_5^{{\rm NLSM}\oplus{\rm Tr}(\phi^3)}(1_\phi,2_\phi,3_\phi,4_\pi,5_\pi)=1-{s_{41}\over s_{451}}-{s_{35}\over s_{345}}\,.~~\label{nlsm+phi-example1-result}
\eea

The result \eref{nlsm+phi-example1-result} obtained via the BCFW method is in complete agreement with that obtained via Feynman rules. This confirms that our claim in section \ref{subsec-key-zero} that such amplitudes have no boundary terms is correct.
Note that the absence of boundary term cannot be established by naive dimensional analysis. This is because each Feynman diagram for ${\cal A}_5^{{\rm NLSM}\oplus{\rm Tr}(\phi^3)}(1_\phi,2_\phi,3_\phi,4_\pi,5_\pi)$ contains a vertex of mass dimension $2$ on the line $L_{(1,3)}$, while the propagator has mass dimension $-2$, so dimensional analysis can only constrain the behavior of ${\cal A}_5(z)$ in the large $z$ limit to be $z/z$.

%%%%%%%%%%%%%%%%%%%%%%%%%%%%%%
\subsection{$5$-point amplitude ${\cal A}_5^{{\rm NLSM}\oplus{\rm Tr}(\phi^3)}(1_\phi,2_\phi,3_\pi,4_\phi,5_\pi)$}
\label{subsec-nlsm+phi-example2}
%%%%%%%%%%%%%%%%%%%%%%%%%%%%%%

The second example is the $5$-point amplitude ${\cal A}_5^{{\rm NLSM}\oplus{\rm Tr}(\phi^3)}(1_\phi,2_\phi,3_\pi,4_\phi,5_\pi)$, which carries three ${\rm Tr}(\phi^3)$ scalars $1\phi$, $2_\phi$, $4_\phi$, and two pions $3_\pi$, $5_\pi$. We choose $i=1$, $j=4$, thus the BCFW-shift in \eref{BCFW-shift} is specialized to
\bea
\hat{k}_1\,=\,k_1+zq\,,~~~~~~\hat{k_4}=k_4-zq\,,~~~~~~~~{\rm where}~k_1\cdot q=k_4\cdot q=q^2=0\,.
\eea
By utilizing the recursive formula \eref{BCFW-recursive-form}, we obtain
\bea
{\cal A}_5^{{\rm NLSM}\oplus{\rm Tr}(\phi^3)}(1_\phi,2_\phi,3_\pi,4_\phi,5_\pi)\,=&&{{\cal A}^{{\rm Tr}(\phi^3)}_3(\hat{1},2,\hat{I})\,{\cal A}^{\rm NLSM}_4(\hat{I},3,\hat{4},5)\Big|_{\hat{s}_{12}(z^*)=0}\over s_{12}}\nn
&+&{\cal B}_5^{{\rm NLSM}\oplus{\rm Tr}(\phi^3)}(\hat{1}_\phi,2_\phi,3_\pi,\hat{4}_\phi,5_\pi)\,.~~~\label{nlsm+phi-2-recur}
\eea
As in the previous example, according to the conclusion obtained in section \ref{subsec-key-zero}, the boundary term ${\cal B}_5^{{\rm NLSM}\oplus{\rm Tr}(\phi^3)}$ is zero. Therefore, the amplitude can be computed entirely from the recursive part in the first line.

The expressions for the two lower-point amplitudes in \eref{nlsm+phi-2-recur} are respectively
\bea
{\cal A}^{{\rm Tr}(\phi^3)}_3(\hat{1},2,\hat{I})\,=\,1\,,~~~~~~~~{\cal A}^{\rm NLSM}_4(\hat{I},3,\hat{4},5)=-s_{35}\,.~~\label{loweramp-example2}
\eea
Here, we have deliberately chosen the expression for ${\cal A}^{\rm NLSM}_4$ so that it is manifestly independent of $z$, thereby simplifying the computation. Since the expressions for ${\cal A}^{{\rm Tr}(\phi^3)}_3$ and ${\cal A}^{\rm NLSM}_4$ are both independent of $z$, we in fact do not need to solve the $z^*$ in \eref{nlsm+phi-2-recur}. Therefore, one can directly substitute \eref{loweramp-example2} into \eref{nlsm+phi-2-recur} to obtain
\bea
{\cal A}_5^{{\rm NLSM}\oplus{\rm Tr}(\phi^3)}(1_\phi,2_\phi,3_\pi,4_\phi,5_\pi)\,=\,{-s_{35}\over s_{12}}\,.~~~~\label{nlsm+phi-example2-result}
\eea

Our result \eref{nlsm+phi-example2-result} coincides with that obtained from Feynman rules, which again verifies the conclusion of no boundary term obtained in section \ref{subsec-key-zero}. As in the previous example, naive dimensional analysis can only constrain the behavior of the amplitude in the large $z$ limit to be $z/z$, which is insufficient to exclude the boundary term.

%%%%%%%%%%%%%%%%%%%%%%%%%%%%%%
\section{Examples: NLSM amplitudes}
\label{sec-example-nlsm}
%%%%%%%%%%%%%%%%%%%%%%%%%%%%%%

In this section, we demonstrate how to compute boundary terms using the recursive formula developed in section \ref{subsec-key-split} by evaluating the $6$-point NLSM amplitude ${\cal A}^{\rm NLSM}_6(1,2,3,4,5,6)$. We compute the recursive part of the amplitude following the standard BCFW method, and then compute the boundary term using the recursive formula \eref{recursionform-boundary}. We present two different procedures for calculating the same amplitude: in subsection \ref{subsec-nlsm-example1}, we perform the BCFW-shift in \eref{BCFW-shift} on external momenta $k_1$ and $k_3$; while in subsection \ref{subsec-nlsm-example2}, we perform the BCFW-shift on external momenta $k_1$ and $k_4$. As will be seen, the complete amplitude obtained from our method agrees exactly with the result obtained from Feynman rules. Finally, in subsection \ref{subsec-comment}, we make some general remarks on the calculational procedures.

In section \ref{subsec-key-split}, we mentioned that, in order to ensure that the integrands in \eref{recursionform-boundary} are expressed in terms of the Mandelstam variables in \eref{M-variable}, there are two alternative approaches. We will use the first in subsection \ref{subsec-nlsm-example1} and the second in subsection \ref{subsec-nlsm-example2}.

%%%%%%%%%%%%%%%%%%%%%%%%%%%%%%
\subsection{$6$-point amplitude ${\cal A}^{\rm NLSM}_6(1,2,3,4,5,6)$, shift $k_1$ and $k_3$}
\label{subsec-nlsm-example1}
%%%%%%%%%%%%%%%%%%%%%%%%%%%%%%

In this subsection, we choose $i=1$, $j=3$. The BCFW-shift \eref{BCFW-shift} becomes
\bea
\hat{k}_1=k_1+zq\,,~~~~~~\hat{k}_3=k_3-zq\,,~~~~~~~~{\rm where}~k_1\cdot q=k_3\cdot q=q^2=0\,.~~\label{BCFW-shift-6pt1}
\eea
The full amplitude can be expressed as
\bea
{\cal A}^{\rm NLSM}_6(1,2,3,4,5,6)\,=\,{\cal R}^{\rm NLSM}_6(\hat{1},2,\hat{3},4,5,6)+{\cal B}^{\rm NLSM}_6(z)(\hat{1},2,\hat{3},4,5,6)\,,~~~~\label{6pt1-full}
\eea
where the recursive part ${\cal R}^{\rm NLSM}_6$ reads
\bea
{\cal R}^{\rm NLSM}_6(\hat{1},2,\hat{3},4,5,6)\,=&&{{\cal A}^{\rm NLSM}_4(6,\hat{1},2,\hat{I})\,
{\cal A}^{\rm NLSM}_4(\hat{I},\hat{3},4,5)\Big|_{\hat{s}_{612}(z^*_1)=0}\over s_{612}}\nn
&+&{{\cal A}^{\rm NLSM}_4(5,6,\hat{1},\hat{I})\,
{\cal A}^{\rm NLSM}_4(\hat{I},2,\hat{3},4)\Big|_{\hat{s}_{561}(z^*_2)=0}\over s_{561}}\,.~~~~\label{6pt1-recur}
\eea

The special $z_1^*$ and $z_2^*$ in \eref{6pt1-recur} can be solved as
\bea
&&\hat{s}_{612}(z_1^*)=0~~\to~~z_1^*=-{s_{612}\over 2K_{26}\cdot q}\,,\nn
&&\hat{s}_{561}(z_2^*)=0~~\to~~z_2^*=-{s_{561}\over 2K_{56}\cdot q}\,,~~~~\label{z1z2-for6pt1}
\eea
where $K_\a$ was defined in \eref{defin-K}.
Meanwhile, the lower-point amplitudes in \eref{6pt1-recur} are given by
\bea
&&{\cal A}^{\rm NLSM}_4(6,\hat{1},2,\hat{I})\,=\,-s_{26}\,,~~~~~~{\cal A}^{\rm NLSM}_4(\hat{I},\hat{3},4,5)\,=\,-\hat{s}_{35}\,,\nn
&&{\cal A}^{\rm NLSM}_4(5,6,\hat{1},\hat{I})\,=\,-\hat{s}_{51}\,,~~~~~~{\cal A}^{\rm NLSM}_4(\hat{I},2,\hat{3},4)=-s_{24}\,.~~~~\label{lower-for6pt1}
\eea
By plugging \eref{z1z2-for6pt1} and \eref{lower-for6pt1} into \eref{6pt1-recur}, we obtain
\bea
{\cal R}^{\rm NLSM}_6(\hat{1},2,\hat{3},4,5,6)\,=\,s_{26}\,\Big({s_{35}\over s_{612}}+{k_5\cdot q\over K_{26}\cdot q}\Big)\,+\,
s_{24}\,\Big({s_{51}\over s_{561}}-{k_5\cdot q\over K_{56}\cdot q}\Big)\,,~~~~\label{6pt1-recur-result}
\eea

We now employ the recursion relation \eref{recursionform-boundary} to compute the boundary term. We choose $t=4$ in
\eref{M-variable}. To ensure that ${\cal B}^{\rm NLSM}_6(z)$ is computed from the integrands expressed in terms of the Mandelstam variables in
\eref{M-variable}, we adopt the first approach mentioned at the end of section
\ref{subsec-key-split}, choosing the expressions for
${\cal A}^{\kappa-{\rm shift}}_{n_1}$ and ${\cal A}^{\kappa'-{\rm shift}}_{n+3-n_1}$ to be in terms of the Mandelstam variables in \eref{M-noz}. There are three choices for $k\in B$, namely $k=4$, $k=5$, and $k=6$. We compute the contributions arising from these three choices one by one.

Let us start with $k=4$. The corresponding $2$-split is
\bea
{\cal A}^{\rm NLSM}_6(1,2,3,4,5,6)\,&\xrightarrow[i=1,j=3,k=4]{\eref{kine-condi-split-scalar}}&\,{\cal J}^{\rm NLSM}_{4}(1,2,3,\kappa)\,\times\,{\cal J}^{{\rm NLSM}\oplus{\rm Tr}(\phi^3)}_5(1_\phi,3_\phi,\kappa'_\phi,5_\pi,6_\pi)\,,~~~~\label{2split-6pt1}
\eea
so that the expressions for the two lower-point on-shell amplitudes can be chosen as
\bea
{\cal A}^{\rm NLSM}_{4}(1,2,3,\kappa)=-s_{2\kappa}\,,~~~~~~~~{\cal A}^{{\rm NLSM}\oplus{\rm Tr}(\phi^3)}_5(1_\phi,3_\phi,\kappa'_\phi,5_\pi,6_\pi)
=1-{s_{51}\over s_{561}}-{s_{\kappa'6}\over s_{\kappa'56}}\,,~~\label{loweramp-6pt1-k=4}
\eea
where the expression \eref{nlsm+phi-example1-result} for $5$-point NLSM$\oplus{\rm Tr}(\phi^3)$ amplitude has been used.
It is easy to see that the operation \eref{s-k} yields
\bea
&&{\cal A}^{\rm NLSM}_{4}(1,2,3,\kappa)=-2k_2\cdot k_{\kappa}\,,\nn
&&{\cal A}^{{\rm NLSM}\oplus{\rm Tr}(\phi^3)}_5(1_\phi,3_\phi,\kappa'_\phi,5_\pi,6_\pi)
=1-{s_{51}\over s_{561}}-{2k_6\cdot k_{\kappa'}\over s_{56}+2K_{56}\cdot k_{\kappa'}}\,,~~\label{loweramp-6pt1-shift-k=4}
\eea
while the shift in \eref{shift-for-offshell-CHY} does not change these two expressions.
In the $2$-split \eref{2split-6pt1}, momentum conservation implies
\bea
k_\kappa=K_{456}\,,~~~~~~~~k_{\kappa'}=K_{24}\,,~~\label{kk'-6pt1-k=4}
\eea
and hence we obtain
\bea
{\cal A}^{\kappa-{\rm shift}}_{4}(1,2,3,\kappa)\,\times\,{\cal A}^{\kappa'-{\rm shift}}_5(1_\phi,3_\phi,\kappa'_\phi,5_\pi,6_\pi)
\,\xrightarrow[i=1,j=3,k=4]{\eref{kine-condi-split-scalar}}\,s_{24}\,\Big({s_{51}\over s_{561}}+{s_{46}\over s_{456}}-1\Big)\,.~~\label{Ak-6pt1-k=4}
\eea
Although we have used $s_{456}$ and $s_{561}$ in the above expression for brevity, if they are expanded as
\bea
s_{456}\,=\,s_{45}+s_{56}+s_{46}\,,~~~~~~s_{561}\,=\,s_{56}+s_{61}+s_{51}\,,
\eea
it is easy to see that all $2$-point Mandelstam variables appearing in the expression \eref{Ak-6pt1-k=4} belong to \eref{M-noz} with $t=4$.
Note that, in obtaining \eref{Ak-6pt1-k=4} by substituting \eref{kk'-6pt1-k=4} into \eref{loweramp-6pt1-shift-k=4}, we have imposed the kinematic condition
\eref{kine-condi-split-scalar}. Performing the BCFW-shift \eref{BCFW-shift-6pt1} on
\eref{Ak-6pt1-k=4}, expanding it in
$z$, and then extracting the constant term, we obtain
\bea
&&{1\over 2\pi i}\,\oint_\infty\,{dz\over z}\,\Big({\cal A}^{\kappa-{\rm shift}}_{4}(\hat{1},2,\hat{3},\kappa)\,\times\,{\cal A}^{\kappa'-{\rm shift}}_5(\hat{1}_\phi,\hat{3}_\phi,\kappa'_\phi,5_\pi,6_\pi)\Big)\Big|^{\eref{kine-condi-split-scalar}}_{\eref{M-variable}}\nn
&=&\,s_{24}\,\Big({s_{46}\over s_{45}+s_{56}+s_{46}}-{k_6\cdot q\over K_{56}\cdot q}\Big)\,.~~~~\label{6pt1-B-k=4}
\eea

Then we turn to the case $k=5$. The corresponding $2$-split is
\bea
{\cal A}^{\rm NLSM}_6(1,2,3,4,5,6)\,&\xrightarrow[i=1,j=3,k=5]{\eref{kine-condi-split-scalar}}&\,{\cal J}^{\rm NLSM}_{4}(1,2,3,\kappa)\,\times\,{\cal J}^{{\rm NLSM}\oplus{\rm Tr}(\phi^3)}_5(1_\phi,3_\phi,4_\pi,\kappa'_\phi,6_\pi)\,,~~~~\label{2split-6pt1-k=5}
\eea
hence the expressions for the lower-point on-shell amplitudes are given as
\bea
{\cal A}^{\rm NLSM}_4(1,2,3,\kappa)=-s_{2\kappa}\,,~~~~~~~~{\cal A}^{{\rm NLSM}\oplus{\rm Tr}(\phi^3)}_5(1_\phi,3_\phi,4_\pi,\kappa'_\phi,6_\pi)
=-{s_{46}\over s_{456}}\,,~~\label{loweramp-6pt1-k=5}
\eea
where the expression \eref{nlsm+phi-example2-result} for $5$-point NLSM$\oplus{\rm Tr}(\phi^3)$ amplitude has been used.
The shift in \eref{s-k} and \eref{shift-for-offshell-CHY} yields
\bea
{\cal A}^{\kappa-{\rm shift}}_4=-2k_2\cdot k_\kappa\,,~~~~~~~~{\cal A}^{\kappa'-{\rm shift}}_5(1_\phi,3_\phi,4_\pi,\kappa'_\phi,6_\pi)\,.
=-{s_{46}\over s_{456}}
\eea
In the $2$-split \eref{2split-6pt1-k=5}, momentum conservation implies
\bea
k_\kappa=K_{456}\,,~~~~~~~~k_{\kappa'}=K_{25}\,,~~\label{kk'-6pt1-k=5}
\eea
therefore we get
\bea
{\cal A}^{\kappa-{\rm shift}}_{4}(1,2,3,\kappa)\,\times\,{\cal A}^{\kappa'-{\rm shift}}_5(1_\phi,3_\phi,4_\pi,\kappa'_\phi,6_\pi)
\,\xrightarrow[i=1,j=3,k=5]{\eref{kine-condi-split-scalar}}\,s_{25}\,{s_{46}\over s_{456}}\,.~~\label{Ak-6pt1-k=5}
\eea
The above result \eref{Ak-6pt1-k=5} is a constant under the BCFW-shift in \eref{BCFW-shift-6pt1}, thus is purely a boundary contribution, namely,
\bea
&&{1\over 2\pi i}\,\oint_\infty\,{dz\over z}\,\Big({\cal A}^{\kappa-{\rm shift}}_{4}(\hat{1},2,\hat{3},\kappa)\,\times\,{\cal A}^{\kappa'-{\rm shift}}_5(\hat{1}_\phi,\hat{3}_\phi,4_\pi,\kappa'_\phi,6_\pi)\Big)\Big|^{\eref{kine-condi-split-scalar}}_{\eref{M-variable}}\nn
&=&\,s_{25}\,{s_{46}\over s_{456}}\,.~~~~\label{6pt1-B-k=5}
\eea

Finally, we consider $k=6$. The associated $2$-split is
\bea
{\cal A}^{\rm NLSM}_6(1,2,3,4,5,6)\,&\xrightarrow[i=1,j=3,k=6]{\eref{kine-condi-split-scalar}}&\,{\cal J}^{\rm NLSM}_{4}(1,2,3,\kappa)\,\times\,{\cal J}^{{\rm NLSM}\oplus{\rm Tr}(\phi^3)}_5(1_\phi,3_\phi,4_\pi,5_\pi,\kappa'_\phi)\,,~~~~\label{2split-6pt1-k=6}
\eea
thus the corresponding lower-point amplitudes are
\bea
{\cal A}^{\rm NLSM}_{4}(1,2,3,\kappa)&=&-s_{2\kappa}\,,\nn
{\cal A}^{{\rm NLSM}\oplus{\rm Tr}(\phi^3)}_5(1_\phi,3_\phi,4_\pi,5_\pi,\kappa'_\phi)
&=&1-{s_{35}\over s_{345}}-{s_{4\kappa'}\over s_{45\kappa'}}\nn
&=&1-{s_{4\kappa'}\over s_{45\kappa'}}+{s_{15}+s_{45}+s_{5\kappa'}\over s_{\kappa'1}}\,,~~\label{loweramp-6pt1-k=6}
\eea
where the transformation of ${\cal A}^{{\rm NLSM}\oplus{\rm Tr}(\phi^3)}_5$ is intended to express it in terms of the desired Mandelstam variables in \eref{M-noz}. The shift in \eref{s-k} and \eref{shift-for-offshell-CHY}
leads to
\bea
{\cal A}^{\kappa-{\rm shift}}_{4}(1,2,3,\kappa)&=&-2k_2\cdot k_\kappa\,,\nn
{\cal A}^{\kappa'-{\rm shift}}_5(1_\phi,3_\phi,4_\pi,5_\pi,\kappa'_\phi)
&=&1-{2k_4\cdot k_{\kappa'}\over s_{45}+2K_{45}\cdot k_{\kappa'}}+{s_{15}+s_{45}+2k_5\cdot k_{\kappa'}\over 2k_{\kappa'}\cdot k_1+k_{\kappa'}^2}\,,~~\label{loweramp-6pt1-shift-k=6}
\eea
where $s_{\kappa'1}$ in \eref{loweramp-6pt1-k=6} is deformed as
\bea
s_{\kappa'1}\,\xrightarrow[]{\eref{s-k}}\,2k_{\kappa'}\cdot k_1\,\xrightarrow[]{\eref{shift-for-offshell-CHY}}\,2k_{\kappa'}\cdot k_1+k_{\kappa'}^2\,.
\eea
In the $2$-split \eref{2split-6pt1-k=6}, the momentum conservation requires
\bea
k_\kappa=K_{456}\,,~~~~~~~~k_{\kappa'}=K_{26}\,,
\eea
therefore
\bea
{\cal A}^{\kappa-{\rm shift}}_{4}(1,2,3,\kappa)\,\times\,{\cal A}^{\kappa'-{\rm shift}}_5(1_\phi,3_\phi,4_\pi,5_\pi,\kappa'_\phi)
\,\xrightarrow[i=1,j=3,k=6]{\eref{kine-condi-split-scalar}}\,s_{26}\,\Big({s_{46}\over s_{456}}-{s_{15}+s_{45}+s_{56}\over s_{12}+s_{16}+s_{26}}-1\Big)\,.~~\label{Ak-6pt1-k=6}
\eea
Performing the BCFW-shift in \eref{BCFW-shift-6pt1} and expanding in $z$, we find
\bea
&&{1\over 2\pi i}\,\oint_\infty\,{dz\over z}\,\Big({\cal A}^{\kappa-{\rm shift}}_{4}(\hat{1},2,\hat{3},\kappa)\,\times\,{\cal A}^{\kappa'-{\rm shift}}_5(\hat{1}_\phi,\hat{3}_\phi,4_\pi,5_\pi,\kappa'_\phi)\Big)\Big|^{\eref{kine-condi-split-scalar}}_{\eref{M-variable}}\nn
&=&\,s_{26}\,\Big({s_{46}\over s_{456}}-{k_5\cdot q\over K_{26}\cdot q}-1\Big)\,.~~~~\label{6pt1-B-k=6}
\eea

Putting \eref{6pt1-B-k=4}, \eref{6pt1-B-k=5} and \eref{6pt1-B-k=6} together, the full boundary term is found to be
\bea
{\cal B}^{\rm NLSM}_6(z)(\hat{1},2,\hat{3},4,5,6)\,=\,s_{24}\,\Big({s_{46}\over s_{456}}-{k_6\cdot q\over K_{56}\cdot q}\Big)+s_{25}\,{s_{46}\over s_{456}}+s_{26}\,\Big({s_{46}\over s_{456}}-{k_5\cdot q\over K_{26}\cdot q}-1\Big)\,.~~~~\label{6pt1-B-result}
\eea
It can be seen that, when the integrands in \eref{recursionform-boundary} are expressed in terms of the Mandelstam variables in \eref{M-variable}, the obtained boundary term indeed satisfies the property analyzed in section \ref{subsec-key-zero}: it can be expanded in the form \eref{expan-doundary}. Each coefficient $C_{ab}$ in the expansion is independent of $s_{a'b'}$, and each choice of $k$ independently probes the subset $\{C_{ak}\}$ of $\{C_{ab}\}$.

By plugging \eref{6pt1-recur-result} and \eref{6pt1-B-result} into \eref{6pt1-full}, we obtain the full amplitude,
\bea
{\cal A}^{\rm NLSM}_6(1,2,3,4,5,6)\,&=&\,{s_{26}s_{35}\over s_{612}}+{s_{24}s_{51}\over s_{561}}+(s_{24}+s_{25}+s_{26})\,{s_{46}\over s_{456}}-s_{24}-s_{26}\nn
&=&\,-{s_{26}(s_{34}+s_{45})\over s_{612}}-{s_{24}(s_{56}+s_{61})\over s_{561}}-{s_{64}(s_{12}+s_{23})\over s_{456}}\nn
&=&\,{(s_{12}+s_{23})(s_{45}+s_{56})\over s_{456}}+{(s_{23}+s_{34})(s_{56}+s_{61})\over s_{561}}+{(s_{34}+s_{45})(s_{61}+s_{12})\over s_{612}}\nn
&&-(s_{12}+s_{23}+s_{34}+s_{45}+s_{56}+s_{61})\,,
\eea
which coincides with the result given by Feynman rules. This indicates that our computation of the boundary term is correct.

%%%%%%%%%%%%%%%%%%%%%%%%%%%%%%
\subsection{$6$-point amplitude ${\cal A}^{\rm NLSM}_6(1,2,3,4,5,6)$, shift $k_1$ and $k_4$}
\label{subsec-nlsm-example2}
%%%%%%%%%%%%%%%%%%%%%%%%%%%%%%

In this subsection, we choose $i=1$, $j=4$. The BCFW-shift in \eref{BCFW-shift} is specialized to
\bea
\hat{k}_1=k_1+zq\,,~~~~~~\hat{k}_4=k_4-zq\,,~~~~~~~~{\rm where}~k_1\cdot q=k_4\cdot q=q^2=0\,.~~~~\label{BCFW-shift-6pt2}
\eea
The complete amplitude can be decomposed as
\bea
{\cal A}^{\rm NLSM}_6(1,2,3,4,5,6)\,=\,{\cal R}^{\rm NLSM}_6(\hat{1},2,3,\hat{4},5,6)+{\cal B}^{\rm NLSM}_6(\hat{1},2,3,\hat{4},5,6)\,,~~~\label{6pt2-full}
\eea
where the recursive part is given as
\bea
{\cal R}^{\rm NLSM}_6(\hat{1},2,3,\hat{4},5,6)\,&=&\,{{\cal A}^{\rm NLSM}_4(6,\hat{1},2,\hat{I})\,{\cal A}^{\rm NLSM}_4(\hat{I},3,\hat{4},5)\Big|_{\hat{s}_{612}(z_1^*)=0}\over s_{612}}\nn
&&+{{\cal A}^{\rm NLSM}_4(5,6,\hat{1},\hat{I})\,{\cal A}^{\rm NLSM}_4(\hat{I},2,3,\hat{4})\Big|_{\hat{s}_{561}(z_2^*)=0}\over s_{561}}\nn
&&+{{\cal A}^{\rm NLSM}_4(\hat{1},2,3,\hat{I})\,{\cal A}^{\rm NLSM}_4(\hat{I},\hat{4},5,6)\Big|_{\hat{s}_{123}(z_3^*)=0}\over s_{123}}\,.~~~\label{6pt2-recur}
\eea

The values of $z_1^*$, $z_2^*$ and $z_3^*$ in \eref{6pt2-recur} can be computed as
\bea
&&\hat{s}_{612}(z_1^*)=0\,\to\,z_1^*=-{s_{612}\over 2K_{26}\cdot q}\,,\nn
&&\hat{s}_{561}(z_2^*)=0\,\to\,z_2^*=-{s_{561}\over 2K_{56}\cdot q}\,,\nn
&&\hat{s}_{123}(z_3^*)=0\,\to\,z_3^*=-{s_{123}\over 2K_{23}\cdot q}\,,~~~~\label{z1z2-for6pt2}
\eea
while the lower-point on-shell amplitudes are given by
\bea
&&{\cal A}^{\rm NLSM}_4(6,\hat{1},2,\hat{I})\,=\,-s_{26}\,,~~~~~~{\cal A}^{\rm NLSM}_4(\hat{I},3,\hat{4},5)\,=\,-s_{35}\,,\nn
&&{\cal A}^{\rm NLSM}_4(5,6,\hat{1},\hat{I})\,=\,-\hat{s}_{51}\,,~~~~~~{\cal A}^{\rm NLSM}_4(\hat{I},2,3,\hat{4})\,=\,-\hat{s}_{24}\,,\nn
&&{\cal A}^{\rm NLSM}_4(\hat{1},2,3,\hat{I})\,=\,-\hat{s}_{13}\,,~~~~~~{\cal A}^{\rm NLSM}_4(\hat{I},\hat{4},5,6)\,=\,-\hat{s}_{46}\,.~~~~\label{lower-for6pt2}
\eea
By substituting \eref{z1z2-for6pt2} and \eref{lower-for6pt2} into \eref{6pt2-recur}, we find the following recursive part
\bea
{\cal R}^{\rm NLSM}_6(\hat{1},2,3,\hat{4},5,6)\,&=&\,{s_{26}s_{35}\over s_{612}}+{s_{13}s_{46}\over s_{123}}+{s_{51}s_{24}\over s_{561}}\nn
&&+s_{13}\,{k_6\cdot q\over K_{23}\cdot q}-s_{46}\,{k_3\cdot q\over K_{23}\cdot q}+s_{51}\,{k_2\cdot q\over K_{56}\cdot q}-s_{24}\,{k_5\cdot q\over K_{56}\cdot q}\nn
&&-s_{123}\,{(k_3\cdot q)\,(k_6\cdot q)\over(K_{23}\cdot q)^2}-s_{561}\,{(k_5\cdot q)\,(k_2\cdot q)\over(K_{56}\cdot q)^2}\,.~~~~\label{6pt2-recur-result}
\eea

Now we compute the boundary term via the recursive formula \eref{recursionform-boundary}. We choose $t=5$ in \eref{M-variable}. This time we adopt the second approach mentioned at the end of section \ref{subsec-key-split}: we start from the arbitrary expressions for ${\cal A}^{\kappa-{\rm shift}}_{n_1}$ and ${\cal A}^{\kappa'-{\rm shift}}_{n+3-n_1}$, and finally replace the Mandelstam variables subject to the constraint from $2$-split condition \eref{kine-condi-split-scalar}, so that the resulting boundary contributions are consistent with the Mandelstam variables in \eref{M-variable}. There are two choices for the external leg $k$, namely $k=5$ and $k=6$. We compute separately the boundary contributions accessible in these two cases.

We first consider $k=5$. The corresponding $2$-split is
\bea
{\cal A}^{\rm NLSM}_6(1,2,3,4,5,6)\,&\xrightarrow[i=1,j=4,k=5]{\eref{kine-condi-split-scalar}}&\,{\cal J}^{{\rm NLSM}\oplus{\rm Tr}(\phi^3)}_{5}(1_\phi,2_\pi,3_\pi,4_\phi,\kappa_\phi)\,\times\,{\cal J}^{\rm NLSM}_4(1,4,\kappa',6)\,,~~~~\label{2split-6pt2-k=5}
\eea
thus the related lower-point on-shell amplitudes are
\bea
{\cal A}^{{\rm NLSM}\oplus{\rm Tr}(\phi^3)}_{5}(1_\phi,2_\pi,3_\pi,4_\phi,\kappa_\phi)=1-{s_{13}\over s_{123}}-{s_{24}\over s_{234}}\,,~~~~~~~~
{\cal A}^{\rm NLSM}_4(1,4,\kappa',6)=-s_{46}\,.~~\label{on1}
\eea
Clearly, neither the shift in \eref{s-k} and \eref{shift-for-offshell-CHY} nor the values of $k_\kappa$ and $k_{\kappa'}$ required by $2$-split affects the above expressions. Thus we immediately get
\bea
{\cal A}^{\kappa-{\rm shift}}_5(1_\phi,2_\pi,3_\pi,4_\phi,\kappa_\phi)\,\times\,{\cal A}^{\kappa'-{\rm shift}}_4(1,4,\kappa',6)
\,\xrightarrow[i=1,j=4,k=5]{\eref{kine-condi-split-scalar}}\,s_{46}\,\Big({s_{13}\over s_{123}}+{s_{24}\over s_{234}}-1\Big)\,.~~\label{Ak-6pt2-k=5}
\eea

Since \eref{Ak-6pt2-k=5} contains the $2$-point Mandelstam variables $s_{24}$ and $s_{34}$ that do not belong to \eref{M-variable}, we need to compute its contribution to the boundary term carefully. Performing the BCFW-shift in \eref{BCFW-shift-6pt2}, \eref{Ak-6pt2-k=5} becomes
\bea
s_{46}\,\Big({s_{13}\over s_{123}}+{s_{24}\over s_{234}}-1\Big)\,\xrightarrow[]{\eref{BCFW-shift-6pt2}}\,(s_{46}-2zk_6\cdot q)\,\Big({s_{13}+2zk_3\cdot q\over s_{123}+2zK_{23}\cdot q}+{s_{24}-2zk_2\cdot q\over s_{234}-2z K_{23}\cdot q}-1\Big)\,.
\eea
Expanding the above expression in z and then extracting the constant term, we obtain
\bea
&&{1\over 2\pi i}\,\oint_\infty\,{dz\over z}\,\Big({\cal A}^{\kappa-{\rm shift}}_5(\hat{1}_\phi,2_\pi,3_\pi,\hat{4}_\phi,\kappa_\phi)\,\times\,{\cal A}^{\kappa'-{\rm shift}}_4(\hat{1},\hat{4},\kappa',6)
\Big)\Big|^{\eref{kine-condi-split-scalar}}_{\eref{M-variable}}\nn
&=&\,(-zk_6\cdot q)\,\Big({s_{13}\over zK_{23}\cdot q}-{s_{24}\over zK_{23}\cdot q}-s_{123}\,{zk_3\cdot q\over z^2(K_{23}\cdot q)^2}+s_{234}\,{zk_2\cdot q\over z^2(K_{23}\cdot q)^2}\Big)\Big|^{\eref{kine-condi-split-scalar}}_{\not{s}_{24},\not{s}_{34}}\nn
&=&\,{k_6\cdot q\over K_{23}\cdot q}\,\Big(s_{24}-s_{13}+s_{123}\,{k_3\cdot q\over K_{23}\cdot q}-s_{234}\,{k_2\cdot q\over K_{23}\cdot q}\Big)\Big|^{\eref{kine-condi-split-scalar}}_{\not{s}_{24},\not{s}_{34}}\nn
&\xrightarrow[i=1,j=4,k=5]{\eref{kine-condi-split-scalar}}&\,{k_6\cdot q\over K_{23}\cdot q}\,\Big(s_{24}-s_{13}+(s_{13}-s_{24}-s_{25})\,{k_3\cdot q\over K_{23}\cdot q}-(s_{24}-s_{13}-s_{35})\,{k_2\cdot q\over K_{23}\cdot q}\Big)\Big|_{\not{s}_{24},\not{s}_{34}}\nn
&=&s_{35}\,{(k_6\cdot q)\,(k_2\cdot q)\over(K_{23}\cdot q)^2}-s_{25}\,{(k_6\cdot q)\,(k_3\cdot q)\over(K_{23}\cdot q)^2}\,.~~~~\label{6pt2-B-k=5}
\eea
As discussed in section \ref{subsec-key-split}, the consistency with the Mandelstam variables in \eref{M-variable} forces us to remove $s_{pj}$, $s_{it}$, and $k_t\cdot q$ from the result of the contour integral around infinity. Therefore, in the above case, what needs to be removed are $s_{24}$ and $s_{34}$.
It is clear that, in the above derivation, the expressions in the third and fifth lines are not equivalent without the constraint of the $2$-split condition \eref{kine-condi-split-scalar}. Therefore, we need to first compute the boundary term consistent with the Mandelstam variables in \eref{M-variable} under the constraint \eref{kine-condi-split-scalar}; otherwise, the result would be incorrect.

We then consider $k=6$. The corresponding $2$-split is given as
\bea
{\cal A}^{\rm NLSM}_6(1,2,3,4,5,6)\,&\xrightarrow[i=1,j=4,k=6]{\eref{kine-condi-split-scalar}}&\,{\cal J}^{{\rm NLSM}\oplus{\rm Tr}(\phi^3)}_{5}(1_\phi,2_\pi,3_\pi,4_\phi,\kappa_\phi)\,\times\,{\cal J}^{\rm NLSM}_4(1,4,5,\kappa')\,,~~~~\label{2split-6pt2-k=6}
\eea
which indicates the following lower-point on-shell amplitudes
\bea
{\cal A}^{{\rm NLSM}\oplus{\rm Tr}(\phi^3)}_{5}(1_\phi,2_\pi,3_\pi,4_\phi,\kappa_\phi)=1-{s_{13}\over s_{123}}-{s_{24}\over s_{234}}\,,~~~~~~~~
{\cal A}^{\rm NLSM}_4(1,4,5,\kappa')=-s_{15}\,.
\eea
Again, neither the shift in \eref{s-k} and \eref{shift-for-offshell-CHY} nor the values of $k_\kappa$ and $k_{\kappa'}$ induced by $2$-split affects the above expressions. Therefore we have
\bea
{\cal A}^{\kappa-{\rm shift}}_5(1_\phi,2_\pi,3_\pi,4_\phi,\kappa_\phi)\,\times\,{\cal A}^{\kappa'-{\rm shift}}_4(1,4,\kappa',6)
\,\xrightarrow[i=1,j=4,k=6]{\eref{kine-condi-split-scalar}}\,s_{15}\,\Big({s_{13}\over s_{123}}+{s_{24}\over s_{234}}-1\Big)\,.~~\label{Ak-6pt2-k=6}
\eea
The BCFW-shift \eref{BCFW-shift-6pt2} turns \eref{Ak-6pt2-k=6} to
\bea
s_{15}\,\Big({s_{13}\over s_{123}}+{s_{24}\over s_{234}}-1\Big)\,\xrightarrow[]{\eref{BCFW-shift-6pt2}}\,(s_{15}+2zk_5\cdot q)\,\Big({s_{13}+2zk_3\cdot q\over s_{123}+2zK_{23}\cdot q}+{s_{24}-2zk_2\cdot q\over s_{234}-2z K_{23}\cdot q}-1\Big)\,.
\eea
Expanding it in $z$ and extracting the constant term, we find
\bea
&&{1\over 2\pi i}\,\oint_\infty\,{dz\over z}\,\Big({\cal A}^{\kappa-{\rm shift}}_5(\hat{1}_\phi,2_\pi,3_\pi,\hat{4}_\phi,\kappa_\phi)\,\times\,{\cal A}^{\kappa'-{\rm shift}}_4(\hat{1},\hat{4},\kappa',6)
\Big)\Big|^{\eref{kine-condi-split-scalar}}_{\eref{M-variable}}\nn
&=&\,(zk_5\cdot q)\,\Big({s_{13}\over zK_{23}\cdot q}-{s_{24}\over zK_{23}\cdot q}-s_{123}\,{zk_3\cdot q\over z^2(K_{23}\cdot q)^2}+s_{234}\,{zk_2\cdot q\over z^2(K_{23}\cdot q)^2}\Big)\Big|^{\eref{kine-condi-split-scalar}}_{\not{s}_{24},\not{s}_{34}}\nn
&=&\,{k_5\cdot q\over K_{23}\cdot q}\,\Big(s_{13}-s_{24}-s_{123}\,{k_3\cdot q\over K_{23}\cdot q}+s_{234}\,{k_2\cdot q\over K_{23}\cdot q}\Big)\Big|^{\eref{kine-condi-split-scalar}}_{\not{s}_{24},\not{s}_{34}}\nn
&\xrightarrow[i=1,j=4,k=6]{\eref{kine-condi-split-scalar}}&\,{k_5\cdot q\over K_{23}\cdot q}\,\Big(s_{13}-s_{24}-(s_{13}-s_{24}-s_{26})\,{k_3\cdot q\over K_{23}\cdot q}+(s_{24}-s_{13}-s_{36})\,{k_2\cdot q\over K_{23}\cdot q}\Big)\Big|_{\not{s}_{24},\not{s}_{34}}\nn
&=&s_{26}\,{(k_5\cdot q)\,(k_3\cdot q)\over(K_{23}\cdot q)^2}-s_{36}\,{(k_5\cdot q)\,(k_2\cdot q)\over(K_{23}\cdot q)^2}\,.~~~~\label{6pt2-B-k=6}
\eea

Putting \eref{6pt2-B-k=5} and \eref{6pt2-B-k=6} together, we obtain the full boundary term
\bea
{\cal B}^{\rm NLSM}_6(\hat{1},2,3,\hat{4},5,6)\,&=&\,s_{35}\,{(k_6\cdot q)\,(k_2\cdot q)\over(K_{23}\cdot q)^2}-s_{25}\,{(k_6\cdot q)\,(k_3\cdot q)\over(K_{23}\cdot q)^2}\nn
&&\,+s_{26}\,{(k_5\cdot q)\,(k_3\cdot q)\over(K_{23}\cdot q)^2}-s_{36}\,{(k_5\cdot q)\,(k_2\cdot q)\over(K_{23}\cdot q)^2}\,.~~\label{6pt2-B-result}
\eea
As can be seen, this boundary term agrees with our conclusion in section \ref{subsec-key-zero}: it can be expanded as \eref{expan-doundary}; each $C_{ab}$ is independent of $s_{a'b'}$, and different choices of $k$ probe different $C_{ak}$ independently. In this example, the choice $k=5$ detects $C_{25}$ and $C_{35}$, whereas the choice $k=6$ detects $C_{26}$ and $C_{36}$.

By substituting \eref{6pt2-recur-result} and \eref{6pt2-B-result} into \eref{6pt2-full}, we finally get
\bea
{\cal A}^{\rm NLSM}_6(1,2,3,4,5,6)\,&=&\,-{s_{26}(s_{34}+s_{45})\over s_{612}}-{s_{24}(s_{56}+s_{61})\over s_{561}}-{s_{64}(s_{12}+s_{23})\over s_{456}}\nn
&&\,+(s_{561}-s_{51}+s_{26}+s_{36}+s_{46})\,{k_2\cdot q\over K_{23}\cdot q}-(s_{123}-s_{13}+s_{24}+s_{25}+s_{26})\,{k_6\cdot q\over K_{23}\cdot q}\nn
&&\,+(s_{123}+s_{561}+s_{25}+s_{35}+s_{26}+s_{36})\,{(k_2\cdot q)\,(k_6\cdot q)\over(K_{23}\cdot q)^2}\nn
&=&\,{(s_{12}+s_{23})(s_{45}+s_{56})\over s_{456}}+{(s_{23}+s_{34})(s_{56}+s_{61})\over s_{561}}+{(s_{34}+s_{45})(s_{61}+s_{12})\over s_{612}}\nn
&&-(s_{12}+s_{23}+s_{34}+s_{45}+s_{56}+s_{61})\,,
\eea
where we have used
\bea
q\cdot K_{56}=-q\cdot K_{1234}=-q\cdot K_{23}\,.
\eea
This result again agrees with the result obtained via Feynman rules. Thus, the boundary term computed in this subsection is also correct.

%%%%%%%%%%%%%%%%%%%%%%%%%%%%%%
\subsection{Remarks}
\label{subsec-comment}
%%%%%%%%%%%%%%%%%%%%%%%%%%%%%%

In this subsection, we make a few remarks on the computations of the two examples presented in previous two subsections.

To ensure that the integrands in the recursive formula \eref{recursionform-boundary} are expressed in terms of the Mandelstam variables in \eref{M-variable}, two alternative approaches were proposed in section \ref{subsec-key-split}. In the first, one expresses ${\cal A}^{\kappa-{\rm shift}}_{n_1}$ and ${\cal A}^{\kappa'-{\rm shift}}_{n+3-n_1}$ in terms of the Mandelstam variables in \eref{M-noz} at the beginning; in the second, one starts from arbitrary expressions for ${\cal A}^{\kappa-{\rm shift}}_{n_1}$ and ${\cal A}^{\kappa'-{\rm shift}}_{n+3-n_1}$, and, subject to the $2$-split condition, deforms the result of the contour integral so as to make it consistent with the Mandelstam variables in \eref{M-variable}. In subsection \ref{subsec-nlsm-example1}, we used the first approach to compute the boundary term for the BCFW-shift \eref{BCFW-shift-6pt1}, while in subsection \ref{subsec-nlsm-example2} we used the second approach to compute the boundary term for the BCFW-shift \eref{BCFW-shift-6pt2}. Clearly, one could equally well use the second approach for the first example, and the first approach for the second example.

For instance, instead of choosing ${\cal A}^{\rm NLSM}_4(1,2,3,\kappa)$ as in \eref{loweramp-6pt1-k=6}, one can also choose
\bea
{\cal A}^{\rm NLSM}_4(1,2,3,\kappa)\,=\,s_{12}+s_{23}\,,
\eea
where $s_{23}$ doe not belong to \eref{M-noz}. A little algebra then leads to
\bea
&&{1\over 2\pi i}\,\oint_\infty\,{dz\over z}\,\Big({\cal A}^{\kappa-{\rm shift}}_{4}(\hat{1},2,\hat{3},\kappa)\,\times\,{\cal A}^{\kappa'-{\rm shift}}_5(\hat{1}_\phi,\hat{3}_\phi,4_\pi,5_\pi,\kappa'_\phi)\Big)\nn
&=&\,(s_{12}+s_{23})\,\Big(1-{s_{46}\over s_{456}}+{k_5\cdot q\over K_{26}\cdot q}\Big)\,.
\eea
To recover the desired formula \eref{6pt1-B-k=6} proportional to $s_{ak}=s_{26}$, one can transform it under the constraint of $2$-split condition, namely,
\bea
s_{12}+s_{23}\,=\,-(s_{24}+s_{25}+s_{26})\,\xrightarrow[]{s_{24}=s_{25}=0}\,-s_{26}\,.
\eea
This is precisely what the second approach does.

As another example, we can recompute the $k=5$ boundary contribution of the second example in \eref{6pt2-B-k=5}, using the first approach.
This time we choose the following expressions for on-shell amplitudes which are equivalent to \eref{on1},
\bea
{\cal A}^{{\rm NLSM}\oplus{\rm Tr}(\phi^3)}_{5}(1_\phi,2_\pi,3_\pi,4_\phi,\kappa_\phi)&=&{s_{13}+s_{3\kappa}\over s_{23}+2K_{23}\cdot K_{1\kappa}}-{s_{13}\over s_{123}}\,,\nn
{\cal A}^{\rm NLSM}_4(1,4,\kappa',6)&=&-s_{1\kappa'}\,,
\eea
then
\bea
{\cal A}^{\kappa-{\rm shift}}_{5}(1_\phi,2_\pi,3_\pi,4_\phi,\kappa_\phi)&=&{s_{13}+2k_3\cdot k_{\kappa}\over s_{23}+2K_{23}\cdot K_{1\kappa}}-{s_{13}\over s_{123}}\,,\nn
{\cal A}^{\kappa'-{\rm shift}}_4(1,4,\kappa',6)&=&-2k_1\cdot k_{\kappa'}-k^2_{\kappa'}\,.
\eea
In the $2$-split \eref{2split-6pt2-k=5}, momentum conservation forces
\bea
k_\kappa\,=\,K_{56}\,,~~~~~~k_{\kappa'}\,=\,K_{235}\,,
\eea
thus
\bea
&&{\cal A}^{\kappa-{\rm shift}}_{5}(1_\phi,2_\pi,3_\pi,4_\phi,\kappa_\phi)\,\times\,{\cal A}^{\kappa'-{\rm shift}}_4(1,4,\kappa',6)\nn
&\xrightarrow[i=1,j=4,k=5]{\eref{kine-condi-split-scalar}}&\,\Big({s_{13}+s_{35}\over s_{23}+2K_{23}\cdot K_{15}}-{s_{13}\over s_{123}}\Big)\,(2K_{1235}\cdot k_6)\,.
\eea
In the above expression, if $s_{123}$, $2K_{23}\cdot K_{15}$ and $2K_{1235}\cdot k_6$ are expanded as
\bea
s_{123}\,&=&\,s_{12}+s_{23}+s_{13}\,,\nn
2K_{23}\cdot K_{15}\,&=&\,s_{12}+s_{13}+s_{25}+s_{35}\,,\nn
2K_{1235}\cdot k_6\,&=&\,s_{16}+s_{26}+s_{36}+s_{56}\,,
\eea
then all Mandelstam variables belong to \eref{M-noz}. Performing the BCFW-shift \eref{BCFW-shift-6pt2} and extracting the constant term, we again find
\bea
&&\,{1\over 2\pi i}\,\oint_\infty\,{dz\over z}\,\Big({\cal A}^{\kappa-{\rm shift}}_5(\hat{1}_\phi,2_\pi,3_\pi,\hat{4}_\phi,\kappa_\phi)\,\times\,{\cal A}^{\kappa'-{\rm shift}}_4(\hat{1},\hat{4},\kappa',6)\nn
&=&\,\Big[{s_{35}\over K_{23}\cdot q}-{k_3\cdot q\over K_{23}\cdot q}\,{s_{23}+2K_{23}\cdot k_5\over K_{23}\cdot q}+{k_3\cdot q\over K_{23}\cdot q}\,{s_{23}\over K_{23}\cdot q}\Big]\,(k_6\cdot q)\nn
&=&\,s_{35}\,{(k_6\cdot q)\,(k_2\cdot q)\over(K_{23}\cdot q)^2}-s_{25}\,{(k_6\cdot q)\,(k_3\cdot q)\over(K_{23}\cdot q)^2}\,.~~\label{redo-example2}
\eea
Note that the derivation from the second line to the third line of \eref{redo-example2} uses only the definitions, and does not use momentum conservation or the on-shell conditions at all. This is as it should be, since when ${\cal A}^{\kappa-{\rm shift}}_5$ and ${\cal A}^{\kappa'-{\rm shift}}_4$ are expressed in terms of the Mandelstam variables in \eref{M-noz}, the resulting boundary term should automatically take the expansion formula \eref{expan-doundary}.

The $2$-splits used in sections \ref{subsec-nlsm-example1} and \ref{subsec-nlsm-example2} correspond to
$|A|=1,|B|=3$ and $|A|=2,|B|=2$, respectively, where $|A|$ and $|B|$ denote the numbers of elements contained in the sets
$A$ and $B$. We did not choose the case $|A|=3,|B|=1$, because then
${\cal J}_{n_1}$ in \eref{2split-scalar} would become the full $6$-point amplitude that needs to be computed, whereas ${\cal J}_{n_2}$ in \eref{2split-scalar} would reduce to the trivial $3$-point
${\rm Tr}(\phi^3)$ current, namely the constant $1$. In this case, the structure in which the higher-point boundary terms are recursively derived from the lower-point on-shell amplitudes no longer exists. More generally, $n_1$ and $n_2$ in \eref{2split-scalar} satisfy $n_1+n_2=n+3$. For the recursive pattern to exist, one must have $n_1<n,n_2<n$, i.e.,
\bea
1\leq|A|\leq n-4\,,~~~~~~~~2\leq|B|\leq n-3\,.~~~~\label{|B|}
\eea

Finally, we make a brief comment on the computational efficiency. Since each choice of
$k$ detects the corresponding $\{C_{ak}\}$, reducing the number of the choices of
$k$ also reduces the number of detections. As the number of candidates for $k$ is exactly
$|B|$, a smaller set $B$ entails fewer detections in the computation of the boundary term. Therefore,
\eref{|B|} tells us that the minimal number of detections corresponds to $|B|=2$. Of course, the number of detections is only one aspect of simplifying the computation, since the complexity of computing the full amplitude also arises from the number of residues that need to be evaluated, and a smaller number of detections in computing the boundary term does not necessarily imply a smaller number of poles for computing the residues. Moreover, the residue at each pole, as well as each detection of the boundary term, may have a different computational complexity. Consequently, the overall computational complexity is difficult to estimate in advance, and an analysis of this issue lies beyond the intended scope of this paper.

%%%%%%%%%%%%%%%%%%%%%%%%%%%%%%
\section{Conclusion and discussion}
\label{sec-conclu}
%%%%%%%%%%%%%%%%%%%%%%%%%%%%%%

In this paper, we study the boundary terms of NLSM$\oplus{\rm Tr}(\phi^3)$ and NLSM amplitudes by making use of the recently discovered hidden zeros and the factorization behaviour near the zeros, referred to as the $2$-split. According to our analysis, for NLSM$\oplus{\rm Tr}(\phi^3)$ amplitudes with three ${\rm Tr}(\phi^3)$ scalar external legs, the boundary terms vanish if the external legs subjected to the BCFW-shift are ${\rm Tr}(\phi^3)$ scalar particles. The boundary terms of NLSM amplitudes, on the other hand, can be expanded in the form \eref{expan-doundary}, with the expansion coefficients independent of any $s_{a'b'}$, where $a'\in A,b'\in B$. Based on this property, we employ the $2$-split behavior to develop a general prescription for recursively constructing the boundary terms of higher-point NLSM amplitudes from lower-point ones. The above conclusions are verified by computing the $5$-point NLSM$\oplus{\rm Tr}(\phi^3)$ amplitudes and the $6$-point NLSM amplitudes. We also compute the $7$-point and $9$-point NLSM$\oplus{\rm Tr}(\phi^3)$ amplitudes with external three ${\rm Tr}(\phi^3)$ scalar legs, as well as the $8$-point and $10$-point NLSM amplitudes. Although the corresponding calculations are too lengthy to be presented here, all the main conclusions in section \ref{sec-fix-boundary} are further verified in these examples.

As mentioned in section \ref{sec-intro}, our ultimate goal is to develop a method for computing boundary terms that covers as many amplitudes satisfying the hidden zeros and $2$-splits as possible. Therefore, the next step is to generalize the method proposed in this paper to amplitudes of other models. The first example is YM amplitudes. Although it was shown in \cite{Arkani-Hamed:2008bsc} that for a given set of external gluon helicities one can always find a suitable BCFW-shift for which no boundary terms arise, we may adopt a different perspective and consider a general formal shift, namely, \eref{BCFW-shift} plus
\bea
\hat{\epsilon}_i\,=\,\epsilon_i-z\,{\epsilon_i\cdot q\over k_i\cdot k_j}\,k_j\,,~~~~~~~~\hat{\epsilon}_j\,=\,\epsilon_j+z\,{\epsilon_j\cdot q\over k_i\cdot k_j}\,k_i\,.
\eea
It is easy to verify that the above BCFW-shift automatically ensures the on-shellness of all external gluons. For such a general shift, the boundary terms cannot be avoided, and a method for computing them is thus required.

When attempting the above generalization, it becomes clear that much of the discussion in section \ref{sec-fix-boundary} relies heavily on the special properties of NLSM$\oplus{\rm Tr}(\phi^3)$ and NLSM amplitudes. When other amplitudes satisfying the hidden zeros and $2$-splits are considered, many of the following properties may be spoiled, including the mass dimensions of the vertices, the absence of $\eta_{\mu\nu}$ in the propagators, the ordering of the amplitudes, and the kinematic conditions for realizing the zeros or $2$-splits (which require additional conditions on the polarization vectors or polarization tensors). On the other hand, the complementary relation between the $2$-splits and the factorizations on the poles observed in section \ref{subsec-0-and-split} still holds, and it continues to suggest that the $2$-splits may compensate for what is missing in the usual BCFW method.

The various new situations that may arise during generalization suggest that, perhaps it is not the best strategy to follow the approach of this paper and analyze each new case theory by theory. It is quite possible to reinterpret the recursive form \eref{recursionform-boundary} for computing boundary terms from a different perspective, such as the CHY formula or surfaceology. Such a new understanding might make the generalization much easier and more systematic. In fact, hidden zeros and $2$-splits have been extensively studied in recent years, and these works contain various ways of understanding the zeros and $2$-splits \cite{Guevara:2024nxd,Cao:2025hio,Zhang:2025zjx,Zhang:2026dcm,Saha:2026ftv,Azevedo:2025vxo,
Feng:2025ofq,Feng:2025dci,Li:2025suo,Backus:2025hpn,Zhou:2026ukg}. With the help of these results, it may be possible for us to achieve a reinterpretation of \eref{recursionform-boundary}. In addition, other zeros and new factorizations near the zeros, which are different from but closely related to the forms given in section \ref{subsec-0-and-split}, have been discovered in the literature \cite{Zhang:2024efe,Li:2026cwa}. These results also provide a new basis for developing methods for computing boundary terms.

%%%%%%%%%%%%%%%%%%%%%%%%%%%%%%%%%%%%%%%%%%%%%%
\section*{Acknowledgments}
%%%%%%%%%%%%%%%%%%%%%%%%%%%%%%%%%%%%%%%%%%%%%%%%%

The author would first like to thank Prof. Bo Feng for his valuable comments and suggestions on the manuscript. The author also thanks Jin Dong, Qu Cao, and Fan Zhu for their answers regarding the expressions for the off-shell currents in the $2$-split. Finally, thanks are due to Prof. L. Rodina and Prof. Yijian Du for discussions on related issues.
This work is supported by NSFC under Grant No. 11805163.

%%%%%%%%%%%%%%%%%%%%%%%%%%%%%%%%%%%%%%%%%%%%%%%%%%%%

%%%%%%%%%%%%%%%%%%%%%%%%%%%%%%%%%%%%%%%%%%%%%%%%%%%%%

\bibliographystyle{JHEP}

\bibliography{reference}

\end{document}